\documentclass[fleqn,usenatbib]{mnras}

\usepackage{newtxtext,newtxmath}
\usepackage[T1]{fontenc}
\usepackage{ae,aecompl}

\usepackage{graphicx}	% Including figure files
\usepackage{amsmath}	% Advanced maths commands
\usepackage{amssymb}	% Extra maths symbols
\usepackage{comment}
\usepackage{longtable}
\usepackage{lscape}
\usepackage{subfigure}
\usepackage{color}
\usepackage{xcolor}

\title[Trends in Massive Star Formation]{Beyond the Solar Circle $-$ Trends in Massive Star Formation Between the Inner and Outer Galaxy}

\author[Djordjevic et al.]{
J. O. Djordjevic,$^{1}$\thanks{E-mail: j.djordjevic@herts.ac.uk}
M.\,A.\,Thompson,$^{1}$
J.\,S.\,Urquhart,$^{2}$
J.\,Forbrich$^{1}$
\\
$^{1}$Centre for Astrophysics Research, University of Hertfordshire, Hatfield AL10 9AB, U.K.\\
$^{2}$School of Physical Sciences, University of Kent, Canterbury CT2 7NZ, U.K.\\
}

\date{Accepted XXX. Received YYY; in original form ZZZ}

\pubyear{2019}

\begin{document}
\label{firstpage}
\pagerange{\pageref{firstpage}--\pageref{lastpage}}
\maketitle
\url

\begin{abstract}
We have compiled the most complete compact and ultracompact H\,{\tiny II} region catalogue to date via multi-wavelength inspection of survey data. We utilise data from the recently available SASSy 850\,$\mu$m survey to identify massive star forming clumps in the outer Galaxy ($R_{\rm{GC}}>8.5$\,kpc) and cross-match with infrared and radio data of known UC H\,{\tiny II} regions from the RMS database. For the inner Galaxy sample ($R_{\rm{GC}}<8.5$\,kpc), we adopt the compact H\,{\tiny II} regions from previous works that used similar methods to cross match ATLASGAL with either CORNISH or RMS, depending on the location within the Galactic plane. We present a new UC H\,{\tiny II} region catalogue that more than doubles the original sample size of previous work, totalling 536 embedded H\,{\tiny II} regions and 445 host clumps. We examine the distance independent values of N$_{\rm{Ly}}/$M and L$_{\rm{bol}}/$M as proxies for massive star formation efficiency and overall star formation efficiency, respectively. We find a significant trend showing that L$_{\rm{bol}}/$M decreases with increasing $R_{\rm{GC}}$, suggesting that the overall star formation per unit mass is less in the outer Galaxy.
\end{abstract}


%%%%%%%%%%%%%%%%%%%%%%%%%%%%%%%%%%%%%%%%%%%%%%%%%%

%%%%%%%%%%%%%%%%% BODY OF PAPER %%%%%%%%%%%%%%%%%%

\vspace{-20mm}
\section{Introduction}

The total observed flux and luminosity of any galaxy is dominated by the massive stars ($M_{*}>8\,\rm{M_{\odot}}$ and $L_{*}>10^3\,\rm{L_{\odot}}$; \citealt{Martins1996}). These stars have a significant impact on their local environments due to strong outflows, stellar winds, optical/UV-radiation, and eventual supernova explosions, affecting the surrounding supply of molecular gas and potentially triggering or quenching any nearby star formation (see review by \citealt{Krumholz2014} and references therein). Galactic evolutionary models that ignore feedback will form stars too early. For this reason, accurate observations of the properties of these massive stars and how they interact with their local environment are needed. The stellar formation process demonstrates the relationship between environment and massive pre-stellar object. The conditions of the local interstellar medium will affect star formation, particularly with regards to metallicity and the gas-to-dust ratio. These factors must be taken into account when deriving and tracing any trends in massive star formation within a host galaxy.

The Milky Way is an ideal landscape for studying the earliest stages of massive star formation. It allows for uniquely high resolution observations of individual star forming regions across the Galactic plane which are often difficult to fully resolve in other nearby galaxies and almost impossible for those with higher redshift, even with the capabilities of modern telescopes such as ALMA \citep{Longmore2014}. The high-resolution data from various Galactic based surveys can provide a template for use in calibrating evolutionary models (e.g. \citealt{Bolatto2008, Kruss2013}). For example, \citealt{Longmore2015} showed that the high-pressure environments of stars in massive stellar clusters such as those found in the central molecular zone are analogous to those of stars forming in galaxies with redshift of $z=1-3$. The key parameters (e.g. gas pressure, surface density, and velocity dispersion) were found to have similar values in both settings. Local conditions will determine the local star formation rates (SFR) and efficiencies (SFE) which can later be applied to extragalactic models to better match observations.

The difficulty lies in accurately tracing the locations of massive star forming sites. The earliest stages of massive star formation occur while the young stellar object (YSO) is still deeply embedded within a dense clump of dust and gas \citep{Garay2004}. These molecular host clumps are typically on the order of a few parsecs across and have masses >$1000$\,M$_{\sun}$ \citep{Solomon1987, Csengeri2014, Urquhart2014agalcsc}. High extinction from the dust makes it difficult to penetrate these thick cocoons \citep{Parsons2009}, but the side effects of star formation are often more readily detectable. For example, methanol masers and compact H\,{\tiny II} regions evolve early on ($<10^5$\,yr; \citealt{Davies2011}) and are easily seen at radio and submillimetre wavelengths. Methanol masers are found to be almost ubiquitously associated with massive star forming clumps and are observed by tracing the 6.7\,GHz emission of massive and highly luminous clouds \citep{Urquhart2013mmbs, Urquhart2014mmbdust}. Compact H\,{\tiny II} regions show up as bubbles of ionised molecular gas within the clump as a direct result of massive star formation occurring within their depths \citep{Wood1989a}. For the purposes of this study, we focus on the compact and ultracompact H\,{\tiny II} regions (hereafter, UC H\,{\tiny II} regions).

Their strong emission in the radio thermal continuum and far-infrared make H\,{\tiny II} regions a valuable tool for tracing Galactic massive star formation \citep{Wood1989a, Kurtz1994, Walsh1998}. An UC H\,{\tiny II} region is formed when a YSO reaches sufficient temperature to emit UV-radiation, ionising the surrounding gas and producing a small region (diameter, $d<0.1$\,pc) of photoionised hydrogen within the host molecular cloud. As the YSO evolves, its UV output increases, expanding the boundaries of the H\,{\tiny II} region so that it may now be defined as compact ($d<0.5$\,pc), classical ($d\sim10$\,pc), or extended ($d>10$\,pc) \citep{Kurtz2005}. A Lyman continuum flux may be derived from the free-free emission (Bremsstrahlung emission) emitted by the hydrogen that directly correlates with the YSO's observed luminosity. Early catalogues of H\,{\tiny II} regions selected candidates based on their mid- or far-infrared colours (e.g. \citealt{Wood1989a, Kurtz1994, Walsh1998}). However, this method also identified other infrared-bright objects such as planetary nebulae and intermediate mass YSOs, leading to confusion in the catalogues and bias in any derived models \citep{Ramesh1997}. The rise of wide area radio surveys prompted a new wave of attempts to separate the similarly-coloured sources through multi-wavelength inspection \citep{Becker1990, Zoonematkermani1990, Becker1994, Giveon2005, Murphy2010, Hindson2012} but many of these were carried out in snapshot mode, limiting UV-coverage and causing some H\,{\tiny II} regions to be falsely identified as bright compact components of more extended emission \citep{Kurtz2000, Kim2001, Ellingsen2005}.

\citealt{Urquhart2013cornish} addressed this problem by cross-matching 870\,$\mu$m submillimetre observations from ATLASGAL with 5\,GHz CORNISH radio data and GLIMPSE mid-infrared 3-colour images (\citealt{Schuller2009,Hoare2012,benjamin2003}, respectively). UC H\,{\tiny II} regions could be recognised from the coincidence of the submillimetre and radio contours embedded in regions of strong dust emission and if they met three primary criteria: (1) they should have radio spectra consistent with thermal free-free continuum, (2) they are clearly associated with thermal infrared emission from the heated dust within the ionised nebula, and (3) they show signs of being embedded within the molecular cloud clump. Within the overlapping area of these surveys ($10\degr \leq l \leq 60\degr$, $\lvert b \vert \leq 1\degr$; see Figure~\ref{fig:survey_coverage}), \citealt{Urquhart2013cornish} were able to identify and confirm 213 bona fide UC H\,{\tiny II} regions embedded within 170 clumps. Kinematic distances were derived for each clump (except where parallax distances were available) and used to estimate clump mass, clump size, Lyman continuum flux, and diameter for each embedded H\,{\tiny II} region. \citealt{Cesaroni2015} performed a similar analysis with Hi-GAL \citep{Molinari2010a} and CORNISH, finding a  comparable number of 230 UC H\,{\tiny II} regions (with $10\degr \leq l \leq 65\degr$, $\lvert b \vert \leq 1\degr$). However, due to the limited $l$ range in CORNISH, these results are incomplete at larger values of galactocentric radii ($R_{\rm{GC}}>8$\,kpc). CORNISH objects within this longitudinal area may also include sources up to a heliocentric distance of D$\sim20$\,kpc, which limits the number of detections that can occur past a certain point and affecting completeness limits (see Figure~\ref{fig:survey_coverage}).

A well-sampled range in $R_{\rm{GC}}$ is essential to fully appreciate the Galactic picture of massive star formation as evidenced by studies such as \citealt{Twarog1997} who analysed a sample of open clusters and found that the Galactic metallicity gradient can be described via two zones, an inner and an outer, separated by a step-like discontinuity around the position of the solar circle ($R_{\rm{GC}} \sim 8.5$\,kpc). Thus, the outer Galactic regions ( $R_{\rm{GC}}> 8.5$\,kpc) will have systematically lower metallicities and higher gas-to-dust ratios compared to the inner Galaxy. \citealt{Lepine2011} and \citealt{Eden2013} further suggested that this variation may be caused by entry shocks of material at the co-rotation radius, confirming such a boundary and a shift in environmental conditions at the position of the solar circle. Later, \citealt{Ness2016} confirmed that metallicity decreases with increasing $R_{\rm{GC}}$, indicating that the material in the outer region is younger and suggesting star formation to be less frequent. Finally, \citealt{Giannetti2017} used observations of the optically thin C$^{18}$O(2-1) transition for 23 massive and dense star-forming regions at $R_{\rm{GC}} > 14$ to determine the approximate gradient by which the gas-to-dust mass ratio, $R$, increased as a function of $R_{\rm{GC}}$ and found the relation, $\log(R) \propto 0.087R_{\rm{GC}}$. This is a significant variation from the model established by \citealt{Draine2007} that assumes a constant value across all galactocentric radii ($R=100$). The results imply that a sample covering the full range in Galactic longitude is necessary to accurately represent the entire Galaxy.

There have been several attempts to compile such a more diverse sample like the all-sky WISE-based one from \citealt{Anderson2014} along with the ongoing Galactic H\,{\tiny II} Region Discovery Survey (HRDS; \citealt{Bania2010}). The catalogue consists of over 8,000 Galactic H\,{\tiny II} regions and H\,{\tiny II} region candidates selected via their characteristic mid-infrared morphologies with the HRDS used to follow up on candidate sources to detect hydrogen radio recombination lines and to confirm their status as a bona fide H\,{\tiny II} region. However, the catalogue focuses on later-stage H\,{\tiny II} regions or those that we consider to be extended or classical regions. HRDS uses the Green Bank Telescope (GBT) which has a FWHM beam size of 82\arcsec \space\space at 9\,GHz (3\,cm). This is too large to accurately sample and identify the UC H\,{\tiny II} regions with expected sizes of approximately 1.5-20\arcsec\ and is not sufficient for our purposes of tracing recent massive star formation via these more compact sources. There are other WISE-based studies that continue to use infrared colour selection methods (e.g. \citealt{Marton2016, Izumi2017}) but remain similarly sensitive to resolution and colour selection issues. Additional attempts that use partial pre-existing H\,{\tiny II} region catalogues (e.g. \citealt{Eden2015, Vut2016}) suffer from the same problems as earlier infrared-based catalogues, including false sources and having large uncertainties. Few of these studies show a sample that provides complete Galactic coverage for larger values of $R_{\rm{GC}}$ and they disagree on the predicted rates by which star formation might decline in the outer Galaxy. \citealt{Wouterloot1988} did push the search for UC H\,{\tiny II} regions to $R_{\rm{GC}} = 14-20$\,kpc using a set of defined far-infrared colour criteria to select likely IRAS candidates and search for $\rm{H_2O}$ masers (another tracer of star formation); however, with a beam size of 4.7\arcmin\space at 100\,$\mu$m, the IRAS catalogue has too much confusion to identify and classify such compact objects with any significant level of confidence \citep{Neugebauer1984}.

Our goal in this paper is to use the \citealt{Urquhart2013cornish} sample of H\,{\tiny II} regions as a starting point to build the rest of a catalogue that will be well-sampled, reliable, and complete for a significant range of $R_{\rm{GC}}$. The SCUBA-2 Ambitious Sky Survey (SASSy; \citealt{Thompson2007}), offers an outer Galaxy submillimetre survey comparable to ATLASGAL in overall angular resolution and sensitivity (see Section 2 for details) but which focuses on Galactic longitudes of $60\degr \leq l \leq 240\degr$. As Figure~\ref{fig:survey_coverage} illustrates, the SASSy data set is able to provide a more complete sample of objects in $R_{\rm{GC}} = 8$-$20$\,kpc than available in ATLASGAL due to the overall closer heliocentric distances for larger larger corresponding values of $R_{GC}$. 

For this work, we adopt the UC H\,{\tiny II} regions found by \citealt{Urquhart2014agalrms} for the remaining ATLASGAL area ($300\degr \leq l \leq 10\degr$) and add in the high angular resolution results of SASSy ($60\degr \leq l \leq 240\degr$) to build upon the results of \citealt{Urquhart2013cornish} and further examine the changes in star formation efficiency as a function of $R_{GC}$. Both new subsets used the same methodology to cross-match the submillimetre with radio and infrared data available from the RMS database which covers the Galactic plane from $10\degr \leq l \leq 350\degr$ \citep{Lumsden2013}. RMS avoids much of the confusion introduced by other infrared-bright objects by filtering the bona fide H\,{\tiny II} regions via follow-up radio observations. The details of these surveys and the resulting radio catalogues are described in Section\,2. The initial \citealt{Urquhart2013cornish} sample was ultimately complemented with an additional 323 bona fide UC H\,{\tiny II} regions. This more than doubles the original sample size, bringing the full number to 536 regions. We explain the detailed source compilation, classification, and matching process used to build the full catalogue in Section 3. In Sections 4 and 5, we present the relevant clump and H\,{\tiny II} region properties and determine whether any variations exist between the inner and outer Galaxy subsets. In Section 6, we examine the resulting trends as a function of $R_{\rm{GC}}$ and compare the findings to other studies that have investigated Galactic trends for  massive SFRs and SFEs. Finally, in Section 7, we summarise our results and discuss the implications.

\section{Surveys}

\begin{figure}
	\includegraphics[width=\columnwidth]{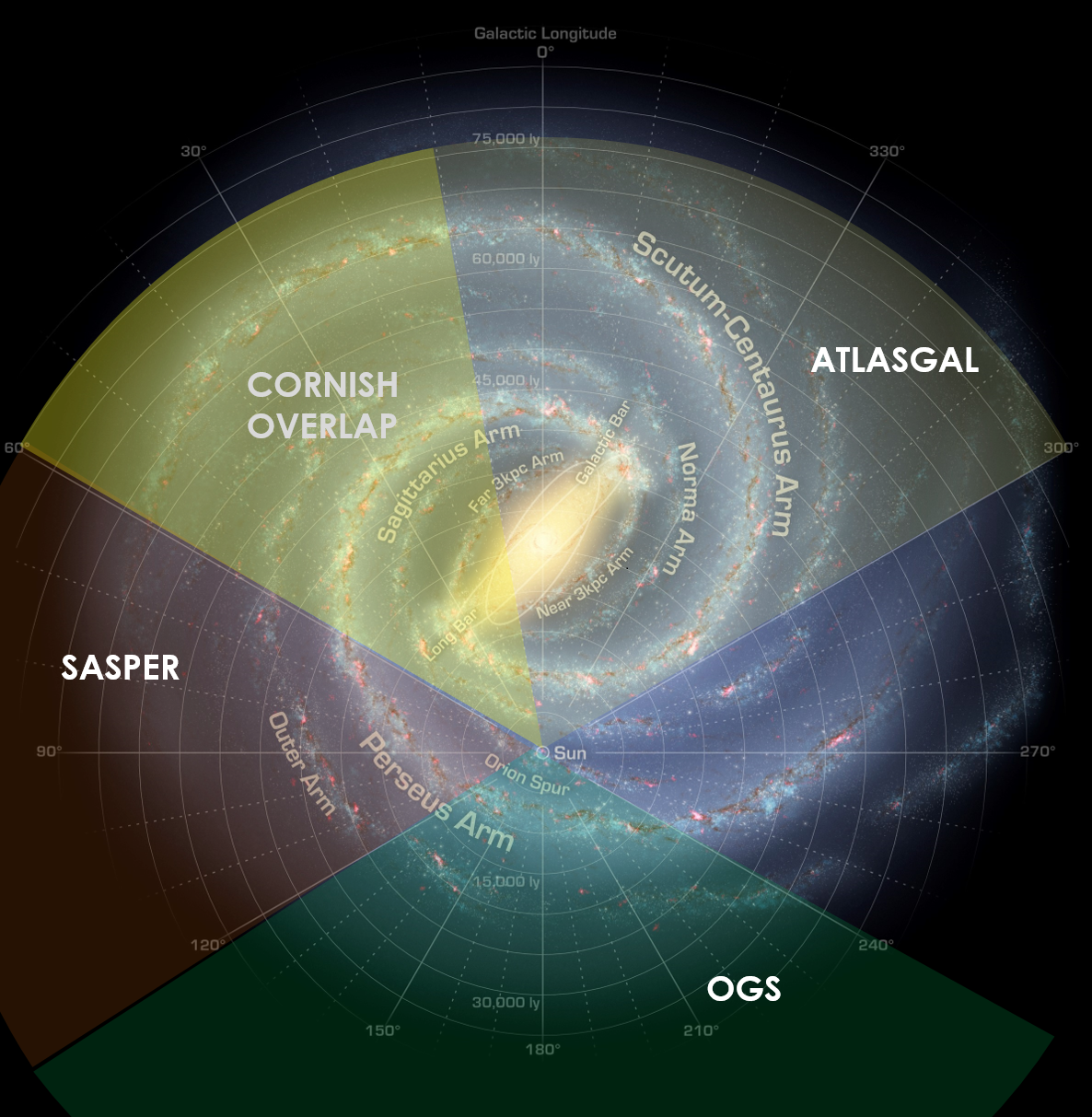}
    \caption{A schematic of the Milky Way showing the Galactic coverage of surveys relevant to this work. ATLASGAL is indicated by the gold shading, whereas the two components of SASSy (SASSy-Perseus or `SASPER' and SASSy-Outer-Galaxy-Survey or `OGS') are shown in red and green, respectively (see text for details). The CORNISH region which overlaps ATLASGAL from $10\degr \leq l \leq 60\degr$ is also shown. RMS covers the full Galactic range from $10\degr \leq l \leq 350\degr$ though it is not shown here so as to avoid cluttering the image. The background image is an artist's impression of the Galactic plane and includes generic points of reference such as larger spiral arms and their names, location of the Sun, and the Galactic bar. To remain consistent with work by \citealt{Urquhart2013cornish} and \citealt{Urquhart2014agalrms}, we adopt the value of $R_{\rm{GC}}=8.5$\, kpc for the distance of the Sun from the Galactic Centre and radius of the solar circle.}
    \label{fig:survey_coverage}
\end{figure}

The following section describes the surveys used in this study. The breakdown in Galactic longitudes for each is shown in Figure~\ref{fig:survey_coverage}.

\subsection{ATLASGAL}

The ATLASGAL survey was the first systematic submillimetre survey of the Galactic plane, covering $300\degr \leq l \leq 60\degr$ and |$b$| $\leq 1.5\degr$, but was later extended to include $280\degr \leq l \leq 300\degr$ with $2\degr \leq b \leq 1\degr$ to account for the warp of the Galactic disk \citep{Schuller2009,Csengeri2014}. The survey used the Large APEX Bolometer Camera  which consists of 295 bolometers observing at 870\,$\mu$m or 345\,GHz \citep{Siringo2009}. For this particular wavelength, the telescope has a beam size of 19.2\arcsec\ at full-width-half-maximum (FWHM) and a positional accuracy of 4\arcsec\ \citep{Schuller2009}. \citealt{Contreras2013} used the source extraction algorithm SExtractor \citep{Bertin1996} to produce an initial compact source catalogue of 6,774 sources for the central region of the survey area ($330\degr \leq l \leq 21\degr$). The full ATLASGAL compact source catalogue (CSC; \citealt{Contreras2013,Urquhart2014agalcsc}) consists of $\sim$10,000 sources and is 97\% complete for sources above 5$\sigma$ and $>99$\% complete above 7$\sigma$ using a normal distribution of noise values and a standard deviation of $\sigma=60$\,mJy beam$^{-1}$. The results from the survey have provided a complete census of dense dust clumps within the inner Galaxy including all potential massive star-forming clumps with masses greater than 1,000\,M$_{\odot}$ out to a heliocentric distance of $\sim$20\,kpc \citep{Urquhart2014agalcsc}.

\subsection{SASSy}

The SCUBA-2 Ambitious Sky Survey (SASSy) utilised the James Clerk Maxwell Telescope (JCMT) to observe the outer Galaxy at 850\,$\mu$m, with an angular resolution of 17\arcsec\ \citep{Dempsey2013}. The results of this survey are currently being prepared for publication (Thompson et al. in prep.). It was designed to fully exploit SCUBA-2's fast mapping capability and to form a long wavelength counterpart to Herschel's PACS and SPIRE \citep{Poglitsch2010,griffin2008}, via the targeting of cold, early-stage objects \citep{Thompson2007}. It covered $\sim$500\,deg$^{2}$ of the sky visible from Mauna Kea down to a 1$\sigma$ noise level of $\sim$30\,mJy beam$^{-1}$. It was the first ground-based submillimetre survey to target specifically the outer Galaxy. As such, it may be used as a beneficial complement to ATLASGAL, possessing similar beam size and sensitivity limits. SASSy was split into two parts: SASSy-Perseus and SASSy-Outer-Galaxy-Survey (hereafter, SASPER and OGS, respectively). SASPER was designed as an extension to the original science verification data when it was announced that Hi-Gal \citep{Molinari2010a} would be increasing its coverage to include the $60\degr \leq l \leq 120\degr$ region, i.e. the Perseus arm (visible in Figure~\ref{fig:survey_coverage}). It was concluded with 100\% completion, covering $\sim$100 deg$^2$ at a nominal depth of 25\,mJy beam$^{-1}$. The catalogue for SASPER consists of 1,372 individual sources (Thompson et al. in prep). OGS covered most of the remaining Galactic regions with $120\degr \leq l \leq 240\degr$. \citealt{Nettke2017} have produced a partial catalogue of 265 beam-sized sources for the $120\degr \leq l \leq 140\degr$ region, down to an rms of $\sim$40\,mJy. The complete OGS catalogue contains 1,766 sources with 99\% completeness above 5$\sigma$ (Thompson et al. in prep.). Altogether, SASSy contains 3,138 clump sources. Both of the subregions in SASSy used the FellWalker source extraction algorithm tool \citep{Berry2014} to produce their respective catalogues. FellWalker was also the method used by JCMT Plane Survey (JPS) which previously used the SCUBA-2 instrument to observe the $7^{\degr} \leq l \leq 63^{\degr}$ portion of the inner Galaxy \citep{Eden2017}.

\subsection{RMS}

The Red MSX Survey (RMS)\footnote{\url{http://rms.leeds.ac.uk/cgi-bin/public/RMS_DATABASE.cgi}} provides a Galactic sample of massive young stellar candidates identified from the MSX satellite catalogue \citep{Price2001} by comparing their mid-infrared colours to those of objects already identified as confirmed MYSOs and H\,{\tiny II} regions \citep{Lumsden2002}. It covers the majority of the Galactic plane with $10\degr \leq l \leq 350\degr$ and |$b$| $\leq 5\degr$, purposely avoiding longitudes near the Galactic centre to circumvent problems with source confusion as a result of crowded positions and/or sources whose kinematic distances cannot be constrained. Though initially selected via these colours, the database has been continuously added to and refined by a programme of multi-wavelength observations, including radio continuum, molecular line, and mid-infrared observations \citep{Hoare2005, Urquhart2007,Urquhart2008rmsMYSOs, Urquhart2008, Urquhart2009, Mottram2011}. A full description of these is provided in \citealt{Lumsden2013}. This process resulted in a catalogue containing MYSOs, H\,{\tiny II} regions, evolved stars, planetary nebulae, and nearby low-mass YSOs. \citealt{Lumsden2013} classified each of these objects according to their 3-colour images provided by GLIMPSE \citep{benjamin2003}, WISE \citep{Wright2010ThePerformance}, or Hi-GAL \citep{Molinari2010a} surveys, as available, and the radio contours from counterparts identified in the follow-up observations. Approximately 1,700 massive YSOs and H\,{\tiny II} regions have been identified to date with radial velocities and distances available for 90\% of the objects \citep{Urquhart2014agalrms}. An analysis from \citealt{Lumsden2013} shows that RMS is more than 90\% complete for the massive protostellar population within the adopted selection boundaries with a positional accuracy of the exciting source of better than 2\arcsec.

\subsection{CORNISH}

The CORNISH survey was used by \citealt{Urquhart2013cornish} as a radio-counterpart to a corresponding area overlapping with ATLASGAL. CORNISH mapped 5\,GHz radio continuum emission in the northern Galactic plane for $10\degr \leq l \leq 60\degr$ and $|b|\leq1$\degr. It was designed to identify UC H\,{\tiny II} regions across the Galactic disk \citep{Hoare2012} and to give a radio counterpart to arcsecond-resolution infrared surveys (e.g. UKIDSS, GLIMPSE, and MIPSGAL). CORNISH used the Very Large Array (VLA) to resolve radio emission on angular scales between 1.5-20\arcsec. The rms noise level of the images is better than 0.4\,mJy beam$^{-1}$, which is sufficient to detect free-free emission from an optically thin H\,{\tiny II} region around a B0 star on the other side of the Galaxy \citep{Urquhart2013cornish}. The CORNISH catalogue contains 2,637 sources above a 7$\sigma$ intensity cut-off \citep{Purcell2013}.

\section{The Catalogue}

The classification of a bona fide UC H\,{\tiny II} region relies on examining the alignment of radio, infrared, and submillimetre data. Several studies have used this technique to classify radio emission, including \citealt{Urquhart2009,Urquhart2013cornish}, \citealt{thompson2006}, \citealt{Hindson2012}, and \citealt{Purcell2013}. The coincidence of the submillimetre data with peaks in the radio and correlation with strong mid-infrared emission has proved a useful tool in confidently identifying objects that are often very similar when viewed via mid-infrared colours alone. 

The compilation of the full catalogue is based on ATLASGAL and SASSy submillimetre datasets with ATLASGAL primarily representing the inner Galaxy and SASSy, the outer Galaxy. In addition to the final source lists, ATLASGAL and SASSy produced clump mask images marking the extent and location of each detected clump source. To ensure consistency between the survey results, we compared their clump-finding methods. As described in Section 2, ATLASGAL identified clumps using the source extraction algorithm SExtractor whereas SASSy used FellWalker. We selected an area around the W3 star-forming complex and ran both algorithms using the SExtractor parameters given in \citealt{Contreras2013}. After excluding sources with fewer than 12 pixels, FellWalker found 46 sources while SExtractor produced 47 sources. The algorithms yielded overall similar results with minor variations in pinpointing the centre of a source and in how they distinguished partially blended sources.

We also examined the RMS database which includes radio and associated infrared data for a large majority of its objects. \citealt{Lumsden2013} had utilised follow-up 5\,GHz radio observations from multiple studies to match each source with either GLIMPSE 3.6, 4.5, and 8.0\,$\mu$m, MIPSGAL 24 and 70\,$\mu$m, WISE 3.4, 4.6, 12, and 22\,$\mu$m, or Hi-GAL 70, 160, 250, 350, 500\,$\mu$m data as available (\citealt{benjamin2003,Carey2009,Wright2010ThePerformance,Molinari2010a}; respectively). Each MYSO candidate was then classified according to a manual inspection of the results. Comments regarding the reasoning for each classification decision were recorded in the database (see footnote link provided in Section 2). The follow-up radio observations were completed with ATCA and the VLA, which was also used in the CORNISH survey. Both CORNISH and the follow-up observations are sensitive to radio emission from angular scales of up to 20\arcsec. RMS currently contains over 5,000 objects across the Galactic plane but only $\sim$900 are classified as H\,{\tiny II} regions which includes both compact and extended sources. 

These radio sources were matched with their corresponding submillimetre host clump using positional matching methods. For ATLASGAL sources, we adopt the matches found by \citealt{Urquhart2013cornish} and \citealt{Urquhart2014agalrms} who took radio positions as given by RMS or CORNISH to match the H\,{\tiny II} regions with corresponding submillimetre clumps whose position and extent had been defined by clump masks generated by SExtractor. Next, archival infrared data was inspected. The 3-colour image of the source overlaid with radio and submillimetre contours could be used to confirm proper coincidence and characteristics of an UC H\,{\tiny II} region. Further details on the process of creating the 3-colour/contour postage stamps is given in \citealt{Urquhart2013cornish} along with examples for a bona fide compact H\,{\tiny II} region as well as other objects commonly mistaken for them when examined solely based on infrared colours. We used these same methods to then match associated RMS H\,{\tiny II} regions with SASSy's FellWalker clumps. 

Of the 900 H\,{\tiny II} regions listed in RMS, 751 are located in the area of the ATLASGAL survey ($280\degr \leq l \leq 60\degr$), with 301 of these lying in the CORNISH region ($10\degr \leq l \leq 60\degr$). \citealt{Urquhart2013cornish} identified 213 UC H\,{\tiny II} regions from ATLASGAL-CORNISH matching. \citealt{Urquhart2014agalrms} matched counterparts between ATLASGAL and RMS sources and found an additional 239 sources. Of these, 49 were located in the CORNISH area that had not been detected previously. We have examined these objects and find that the majority appear to have peak fluxes below the CORNISH sensitivity limit of $\sim$2.7\,mJy. Others may have only been detected later as a result of flux variability. \citealt{Kalcheva2018} has estimated that $\sim$5\% of all H\,{\tiny II} regions are affected by this trait. The remaining few sources ($\sim$5) have flux differences of only a few mJy from the CORNISH sensitivity limits which may be a result of differences between data reduction and/or calibration techniques among the various surveys. Lastly, in the SASSy region ($60\degr \leq l \leq 240\degr$), RMS lists 124 H\,{\tiny II} regions and we identified 84 matches with a SASSy counterpart. Overall, the final catalogue consists of 536 H\,{\tiny II} regions associated with 445 molecular host clumps covering the Galactic plane with $280\degr \leq l \leq 240\degr$ and more than doubling the initial sample size from \citealt{Urquhart2013cornish}. We note that there are a relatively small number of matches identified for the SASSy-OGS region but further examination showed that very few of the potential RMS matches in this area had existing radio data. We will address the possible implications of this on the resulting galactocentric radial trends later in Section 6.

We are not concerned with the possibility of chance alignments of Galactic dust emission with extragalactic background sources, planetary nebulae, or radio emission from more extended H\,{\tiny II} regions as these objects were already re-classified as non-H\,{\tiny II} regions and excluded during the RMS classifications made using  follow-up observations.

\begin{figure}
	\includegraphics[width=\columnwidth]{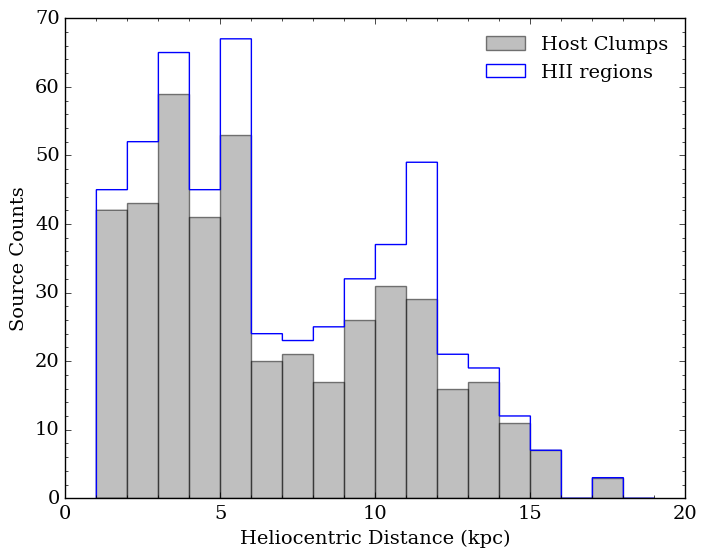}
    \includegraphics[width=\columnwidth]{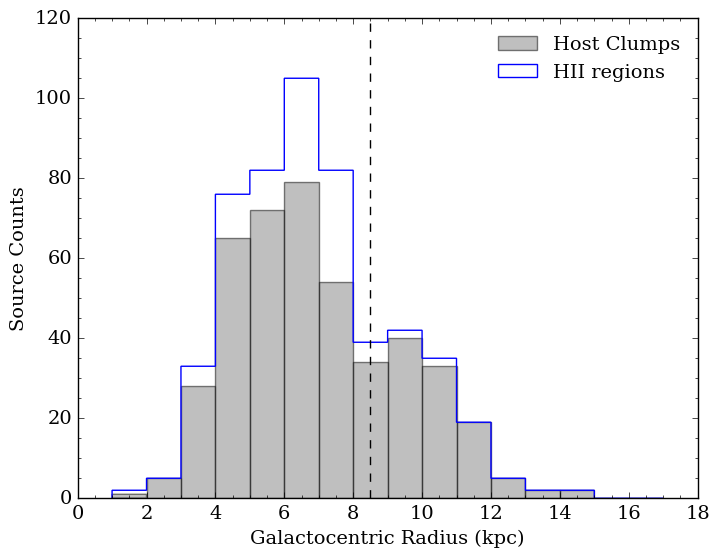}
    \caption{Distance distributions of host clumps (grey) and their embedded UC H\,{\tiny II} regions (blue). The upper panel depicts heliocentric distance and shows a relatively complete distribution of distances between 1 and 20\,kpc away. The lower panel presents the distribution for galactocentric radius. The vertical dashed line at $R_{\rm{GC}}=8.5$\,kpc represents the location of the sun and the corresponding solar circle which we will use as the boundary between the inner Galaxy and the outer Galaxy. On average, the outer Galaxy will have lower metallicities and higher dust-to-gas ratios when compared to the inner Galaxy.}
    \label{fig:dist_histo}
\end{figure}

\subsection{Distances}

Reliable distance estimates are essential for examining the physical properties of the UC H\,{\tiny II} regions and their host clumps. Maser parallax and spectroscopic distances are the preferred methods as these tend to be more reliable. In our catalogue, we found 20 sources known to be associated with the Cygnus X region located at $D=1.4$\,kpc and $R_{\rm{GC}}=8.2$\,kpc for which well known parallax information is available and thus their associated uncertainties will be very small compared to the rest of the sample \citep{Rygl2012}. 

The remaining sources in our catalogue have kinematic distances. It is important to stress we have chosen to use the updated distance and galactocentric radii values provided by \citealt{Urquhart2018} which recalculated kinematic distances for many ATLASGAL clump sources (see distance reference flags included in Table 1). Despite the corrections, the radial velocity measurements tend to have a general uncertainty of $\pm10$\,km $\rm{s}^{-1}$ due to systematic errors from streaming motions \citep{Reid2009}. This corresponds to a kinematic distance error of $\sim$0.6-1\,kpc. For the SASSy sources, we adopt the distance listed for the corresponding radio source in  the RMS catalogue which draws on various references and follow-up studies to provide values for each source (see \citealt{Urquhart2007a, Urquhart2008, Urquhart2014agalrms} and references therein). These values are subject to similar overall distance errors as those found in ATLASGAL. Of the 445 total host clumps, we were unable to determine distances for 6 of them (correlating with 7 H\,{\tiny II} regions) due to radial velocities not being available in RMS or other literature. 

 Figure~\ref{fig:dist_histo} shows the complete distribution of distances for both the host massive star forming clumps (grey) and UC H\,{\tiny II} regions (blue). The upper panel shows the distribution in terms of heliocentric distances with the sample of star forming regions possessing distances ranging from $\sim1$ to $20$\,kpc. There are evident peaks in massive star forming clumps near 5 and 10\,kpc. We also note several likely peaks near 2.5, 7, and 15\,kpc where there is a higher density of UC H\,{\tiny II} regions present per clumps. \citealt{Urquhart2013cornish} pointed out that these distances roughly coincide with the tangent points of the Galactic spiral arms and the distance between the peaks is larger than the assumed distance errors of $\pm1$\,kpc which also describes the approximate average width of a spiral arm. In the lower panel of Figure~\ref{fig:dist_histo}, the sources are re-plotted in terms of galactocentric radius. The location of the sun at R$_{\rm{GC}} = 8.5$\,kpc is illustrated as a vertical dashed line. The sample is dominated by inner Galaxy sources ($\sim$75\%) with 334 clumps and 420 H\,{\tiny II} regions having $R_{\rm{GC}}\leq8.5$\,kpc. In the outer Galaxy, the catalogue includes 105 clumps and 109 H\,{\tiny II} regions. Again we see potential peaks around 6 and 9\,kpc which correspond to spiral arms positions along our line of sight directly towards and away from the Galactic Centre (refer back to Figure~\ref{fig:survey_coverage}). 
 In the following sections, we will examine how the properties of these clumps and their embedded UC H\,{\tiny II} regions vary as a function of galactocentric radius. We will investigate any trends that appear, in particular how SFRs and SFEs are affected when viewing the relatively denser and higher metallicity inner Galaxy versus the sparser, lower metallicity regions of the outer Galaxy. The inclusion of SASSy data ensures that the sample is investigated out to a Galactocentric radius of 12\,kpc with additional clumps scattered out to 15\,kpc or up to nearly twice the solar radius.

\subsection{Dust Temperatures}

The clump properties for sources adopted from \citealt{Urquhart2013cornish} and \citealt{Urquhart2014agalrms} were initially calculated using a uniform dust temperature of $T_{\rm{dust}}=20$\,K across the Galactic plane. However, \citealt{Urquhart2018} also obtained individual dust temperatures for each ATLASGAL source, showing that temperatures tend to increase by $\sim$5\,K with galactocentric radius between 5 and 15\,kpc. However, no dust temperatures are currently available for the SASSy clumps. For consistency, we have adopted the average dust temperature from those calculated by \citealt{Urquhart2018} and find a mean value of $T_{\rm{dust}}=27\,$K. We will use this value to calculate all corresponding clump and H\,{\tiny II} region properties as presented in Sections 4 and 5, with the caveat that if the temperature should actually increase as a function of radius, then we may be overestimating the masses for clumps at higher galactocentric radii.

\begin{table*}
\centering
\caption{Derived clump properties of the host molecular clumps. The first column gives Submillimetre Name (\textit{AGAL} for ATLASGAL objects; \textit{JCMTLSY} for SASSy) with superscripts denoting publication clump source was adopted from: $^{1}$\citealt{Urquhart2013cornish}; $^{2}$\citealt{Urquhart2014agalrms}; $^{3}$Thompson et. al. in prep. The remaining columns include: Complex that clump belongs to (if any); H\,{\tiny II} Region Density; Radial Velocity; Heliocentric Distance; Galactocentric Radius; Flag denoting source of adopted distance information ([1]\citealt{Urquhart2018}; [2]\citealt{Urquhart2013cornish}; [3]\citealt{Urquhart2014agalrms}; [4]Adopted from RMS database); Effective Radius of clump; Peak and Integrated submillimetre fluxes; Gas-to-dust Ratio Value with \citealt{Giannetti2015} correction; Column Density; Corrected Column Density; Clump Mass; Corrected Clump Mass. Full table will be available in electronic form at CDS.}
\resizebox{\textwidth}{!}{
\label{table1}
\begin{tabular}{lcccccccccccccc}
\hline
\hline
{Submm Name} & {Complex} & {H{\tiny{II}}} & {v$_{\rm{lsr}}$} & {D} & {R$_{\rm{GC}}$} & {Flag} & {Radius} & {Peak Flux} & {Int Flux} & {\textit{R}} & {Log $N_{H_{2}}$} & {Log $N_{H_{2}}$ [corr]} & {Log M} & {Log M [corr]}\\
{ } & { } & { } & {[km\,s$^{-1}$]} & {[kpc]} & {[kpc]} & { } & {[pc]} & {[Jy\,beam$^{-1}$]} & {[Jy]} & { } & {[cm$^{-2}$]} & {[cm$^{-2}$]} & {[M$_{\odot}$]} & {[M$_{\odot}$]}\\
\hline
 $^{1}$AGAL010.299$-$00.147	&	W31-North	&	1	&	12.8	&	3.5	&	4.9	&	1	&	2.77	&	7.67	&	54.18	&	73.5	&	19.13	&	19	&	3.38	&	3.25	\\
 $^{1}$AGAL010.321$-$00.257	&	W31-South	&	1	&	32.2	&	3	&	5.4	&	1	&	2.48	&	2.39	&	14.93	&	81.25	&	18.62	&	18.53	&	2.69	&	2.6	\\
 $^{1}$AGAL010.472+00.027	&	$-$	&	2	&	66.7	&	8.5	&	1.6	&	1	&	5.54	&	35.01	&	88.12	&	37.95	&	19.79	&	19.37	&	4.36	&	3.94	\\
 $^{1}$AGAL037.819+00.412	&	$-$	&	2	&	18	&	12.3	&	7.7	&	1	&	3.58	&	2.94	&	7.57	&	128.8	&	18.71	&	18.82	&	3.62	&	3.73	\\
 $^{1}$AGAL037.867$-$00.601	&	$-$	&	1	&	50.7	&	10	&	6.2	&	2	&	1	&	1.67	&	5.05	&	95.37	&	18.47	&	18.45	&	3.26	&	3.24	\\
 $^{1}$AGAL037.874$-$00.399	&	$-$	&	1	&	60.8	&	9.7	&	6	&	1	&	4.4	&	5.37	&	18.45	&	91.62	&	18.97	&	18.94	&	3.8	&	3.76	\\
 $^{1}$AGAL038.646$-$00.226	&	$-$	&	1	&	69.2	&	4.7	&	5.7	&	2	&	0.53	&	0.84	&	2.86	&	86.28	&	18.17	&	18.11	&	2.36	&	2.3	\\
 $^{1}$AGAL038.652+00.087	&	$-$	&	1	&	-36.5	&	14.6	&	9.6	&	1	&	2.38	&	0.77	&	3.62	&	188.45	&	18.13	&	18.41	&	3.45	&	3.72	\\

 $^{2}$AGAL346.076$-$00.056	&	$-$	&	1	&	-83.9	&	10.2	&	2.9	&	1	&	5.1	&	2.38	&	11.57	&	49.24	&	18.62	&	18.31	&	3.64	&	3.33	\\
 $^{2}$AGAL346.232$-$00.321	&	$-$	&	1	&	-11.5	&	14.98	&	7.02	&	3	&	1.31	&	0.78	&	2.65	&	112.39	&	18.14	&	18.19	&	3.33	&	3.39	\\
 $^{2}$AGAL347.304+00.014	&	$-$	&	1	&	-8.9	&	1.4	&	7	&	1	&	0.16	&	0.69	&	1.68	&	111.94	&	18.08	&	18.13	&	1.08	&	1.13	\\
 $^{3}$JCMTLSY J073538.53$-$184855.1	&	$-$	&	1	&	46.8	&	3.5	&	10.8	&	4	&	0.95	&	0.27	&	15.38	&	239.66	&	17.7	&	18.08	&	2.81	&	3.19	\\
 $^{3}$JCMTLSY J074451.83$-$240744.3	&	$-$	&	1	&	66.8	&	5.4	&	12	&	4	&	2.33	&	2.52	&	179.97	&	304.79	&	18.67	&	19.16	&	4.26	&	4.74	\\
 $^{3}$JCMTLSY J194815.31+280727.4	&	$-$	&	1	&	-55.3	&	11.7	&	11	&	4	&	2.19	&	0.54	&	0.53	&	249.46	&	18	&	18.4	&	2.4	&	2.8	\\
 $^{3}$JCMTLSY J194914.48+265010.9	&	$-$	&	1	&	$-$	&	$-$	&	$-$	&	$-$	&	$-$	&	0.68	&	2.39	&	$-$	&	18.1	&	$-$	&	$-$	&	$-$	\\
 $^{3}$JCMTLSY J195803.01+314407.3	&	$-$	&	1	&	-65.5	&	11.7	&	11.6	&	4	&	2.08	&	0.33	&	0.32	&	281.32	&	17.79	&	18.24	&	2.18	&	2.63	\\
 $^{3}$JCMTLSY J200137.46+333527.5	&	$-$	&	1	&	-22.9	&	7.4	&	9.1	&	4	&	3.75	&	0.54	&	2.28	&	170.49	&	18	&	18.24	&	2.64	&	2.87	\\
 $^{3}$JCMTLSY J200145.71+333244.3	&	$-$	&	1	&	-25.2	&	7.6	&	9.2	&	4	&	7.8	&	12.68	&	47.43	&	173.94	&	19.37	&	19.62	&	3.98	&	4.22	\\

 $^{3}$JCMTLSY J203900.97+421931.5	&	$-$	&	2	&	-2.4	&	1.2	&	8.3	&	4	&	1.52	&	18.33	&	143.38	&	145.24	&	19.53	&	19.7	&	2.85	&	3.02	\\
 $^{3}$JCMTLSY J203901.27+422203.6	&	Cygnus X	&	1	&	-3.8	&	1.4	&	8.3	&	4	&	1.09	&	10.32	&	50.6	&	145.24	&	19.29	&	19.45	&	2.54	&	2.7	\\
 $^{3}$JCMTLSY J203925.53+411959.2	&	Cygnus X	&	1	&	-2	&	1.4	&	8.3	&	4	&	0.81	&	1.08	&	7.28	&	145.24	&	18.31	&	18.47	&	1.69	&	1.86	\\
 $^{3}$JCMTLSY J204233.19+425645.6	&	Cygnus X	&	1	&	-4.1	&	1.4	&	8.3	&	4	&	0.57	&	1.05	&	3.33	&	145.24	&	18.29	&	18.46	&	1.35	&	1.52	\\
 $^{3}$JCMTLSY J205413.82+445408.8	&	$-$	&	1	&	-35.8	&	5.5	&	9.6	&	4	&	2.82	&	1.88	&	6.02	&	188.45	&	18.55	&	18.82	&	2.8	&	3.07	\\

\hline
\end{tabular}
}
\end{table*}

\section{Clump Properties}

\begin{figure}
    \includegraphics[width=\columnwidth]{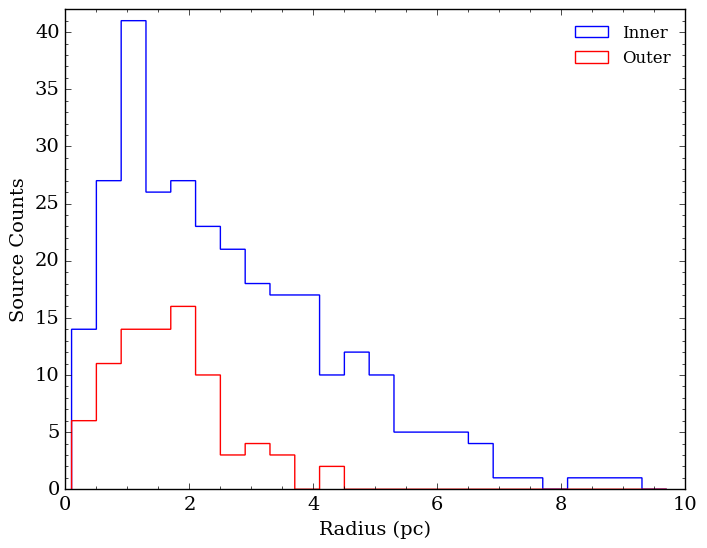}
    \caption{The distribution of inner (blue) vs outer (red) Galaxy sources of effective radii for each clump associated with at least one H\,{\tiny II} region. See text for further details. Bin size is 0.4\,pc.}
    \label{fig:clump_radius_histo}
\end{figure}

In the following section, we present the physical properties of the molecular clumps which host the sample of UC H\,{\tiny II} regions. All calculations are based on the distances and temperature assumption described previously in Section 3. The clump properties for a selection of sources are listed in Table 1 with the full catalogue available online. The results are also summarised with a series of histograms displaying the data as two subsets representing inner and outer Galaxy sources. The inner Galaxy is normally defined as those sources that have $R_{\rm{GC}} < 8.5$ and outer Galaxy sources will have $R_{\rm{GC}}$ above this value. However, due to the $\pm$1\,kpc distance error, we exclude sources within the range of uncertainty. Thus, the figures below display the definite inner Galaxy sources with $2 \leq R_{\rm{GC}} < 7.5$\,kpc while definite outer Galaxy sources have $R_{\rm{GC}} > 9.5$\,kpc. We also exclude 6 individual clump sources whose distances could not be obtained. The host molecular clumps are likely to be forming protoclusters of YSOs. The most massive members of these protoclusters will be responsible for the majority of the observed bolometric luminosity and Lyman continuum flux.

\subsection{Clump Radius}

Effective radii were derived from source areas observed by corresponding submillimetre data. The approximation as originally presented by \citealt{Rosolowsky2010} is

\begin{equation}
          R_{eff} = \sqrt{A/\pi}
\end{equation}

\noindent where A is the area of the clump. For sources with a known distance, the results were converted to a physical size. The resulting distributions are shown in Figure~\ref{fig:clump_radius_histo} for 289 inner Galaxy sources (blue) and 83 outer Galaxy sources (red). Overall, the subsets are similar with peaks between 1-2\,pc. However, the larger inner Galactic sample covers a wider range of radii with sources falling between 0.13 and 12.09\,pc while the outer sources are found in the more limited 0.3-4.37\,pc range. The average inner Galaxy clump size is 2.7\,pc with a median value of 2.3\,pc. The average outer Galaxy clump is smaller at 1.67\,pc and a median of 1.58\,pc. These differences are confirmed via a Kolmogorov-Smirnov test (KS-test) which gives a p-value $\ll 0.001$. Thus, we are able to confidently (within 3$\sigma$) reject the hypothesis that the two subsets are drawn from the same parent sample.

\subsection{Clump Mass}

\begin{figure}
   \includegraphics[width=\columnwidth]{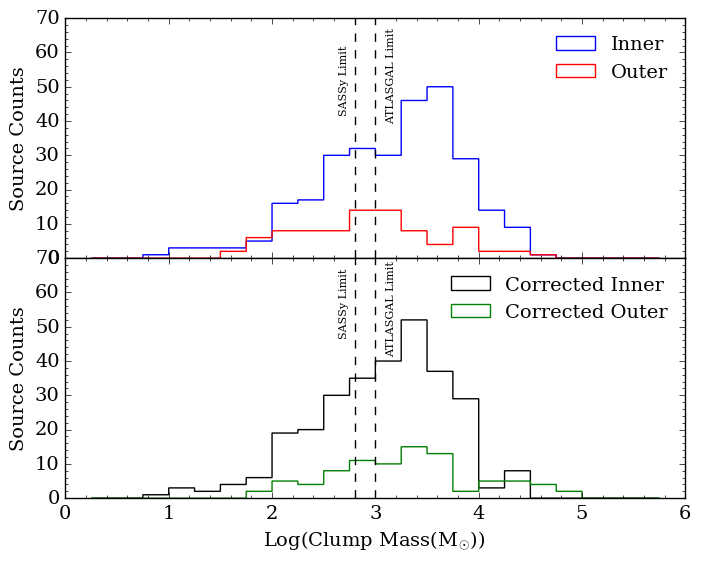}
  \caption{Upper panel: Distribution of clump masses for inner and outer Galaxy sources assuming $R=100$. Lower panel: Distribution of clump mass after corrections have been made for $R$ with respect to each source's $R_{\rm{GC}}$. The mass limits for ATLASGAL and SASSy are depicted via the dashed horizontal lines at 10$^{3}$ and 10$^{2.8}$ M$_{\odot}$, respectively. Bin size is 0.25\,dex. }
  \label{fig:clump_mass}
\end{figure}

The clump mass values were estimated using the methods of \citealt{Hildebrand1983} which assumes that the submillimetre emission is optically thin and represented by a single dust temperature. Overall, the Planck function is nonlinear which can contribute an error of a factor of 2 or 3 in the derived values. We must also recall our assumption that the Galactic disk may be approximated by a uniform dust temperature (see Section 3.2) which may result in us overestimating the clump masses. Assuming that the total clump mass is proportional to the total flux density integrated over the source, the equation is

\begin{equation}
\Big(\frac{M_{\rm{clump}}}{\rm{M_{\odot}}}\Big) = \Big(\frac{D}{\rm{kpc}}\Big)^{2} \Big(\frac{S_{\nu}}{\rm{mJy}}\Big)\frac{R}{B_{\nu}(T_{\rm{dust}})\kappa_{\nu}}
\end{equation}

\noindent where $S_{\nu}$ is the integrated submillimetre flux, $D$ is the heliocentric distance, $R$ is the gas-to-dust mass ratio assumed to be 100 \citep{Draine2007}, $B_{\nu}$ is the Planck function (with $T_{\rm{dust}}=27$\,K; see Section 3), and $\kappa_{\nu}$ is the dust absorption coefficient. For the ATLASGAL sources, $\kappa_{\nu}$ is taken as 1.85 cm$^{2}$ g$^{-1}$ as derived by \citealt{Schuller2009} by interpolating 870 $\mu$m from table 1, column 5 of \citealt{Ossenkopf1994}. Similarly, we use a value of 1.87 cm$^{2}$ g$^{-1}$ for the SASSy 850 $\mu$m sources.

The results are shown in the top panel of Figure~\ref{fig:clump_mass}. However, it is unlikely this gas-to-dust ratio remains constant throughout the Galaxy. In fact, in the outer Galaxy, the metallicity and average disk surface density begin to substantially decrease (e.g. \citealt{Elia2013, Elia2017}; \citealt{Konig2017}). Observations of the Galactic chemical evolution of dust also suggest that $R$ should increase with decreasing metallicity \citep{Dwek1998,Mattsson2012,Hirashita2017} and the same has been observed for the disks of nearby galaxies \citep{Sandstrom2013}. \citealt{Giannetti2017} provided the most comprehensive study of this relationship for a wide range of galactocentric radii and calculated that the gas-to-dust ratio was dependent on radius, increasing via\footnote{In the original paper, Giannetti's equation uses $\gamma$ instead of $R$. We have rewritten the formula here to remain consistent with our nomenclature.}:

\begin{equation}
\rm{log}(\it{R})=\rm{\Big(0.087^{+0.045}_{-0.025}\pm0.007\Big)}\it{R}_{\rm{GC}}+\rm{\Big(1.44^{-0.45}_{+0.21}\pm0.03\Big)}
\end{equation}

In the lower panel of Figure~\ref{fig:clump_mass}, we have calculated ATLASGAL and SASSy clump masses using the corrections given by equation (3). Both values are included in Table 1. The completeness limits for ATLASGAL and SASSy are also shown at $10^{3}$\,M$_{\odot}$ and $10^{2.8}$\,M$_{\odot}$, respectively. The ATLASGAL limit was adopted from \citealt{Urquhart2013cornish} while the representative value for SASSy was estimated by measuring the background rms value of each image and averaging together to represent a flux uncertainty for the full SASSy region. This value was inserted into equation (2) along with the maximum observed distance (8\,kpc) of a matched SASSy-RMS source to derive the corresponding mass completeness limit. 

In the inner and outer Galaxy subsets, there are 289 and 86 clumps, respectively. When the gas-to-dust corrections are applied to the mass calculations, we notice an overall shift towards lower values for the inner Galaxy subset and an opposite shift apparent in the outer Galaxy sample. Inner galaxy uncorrected masses peak around 3.8\,dex with an average value of 3.14\,dex and a median of 3.27\,dex. These are distinctly different from the outer Galaxy values of 3, 2.92, 2.93\,dex, respectively. With the corrected masses, both subsets show a peak around 3.5\,dex. Similarly, the inner and outer Galaxy averages (3.06\,dex vs. 3.3\,dex) and median values (3.18\,dex vs. 3.47\,dex) also show the relative shifts of each subset. 

A KS-test shows that while the inner vs outer Galaxy uncorrected masses have a p-value of $0.012$, the corrected values possess an even lower p-value of $0.006$, This result suggests that there is a $<1$\,\% chance that our rejection of the null hypothesis is incorrect. However it remains outside the established 3$\sigma$ confidence level. In Section 5, we will re-examine the clump masses as they correlate with other parameters of the star forming clumps and embedded UC H\,{\tiny II} regions to better cancel out corresponding sources of error. 

\subsection{Column Density}

\begin{figure}
    \includegraphics[width=\columnwidth]{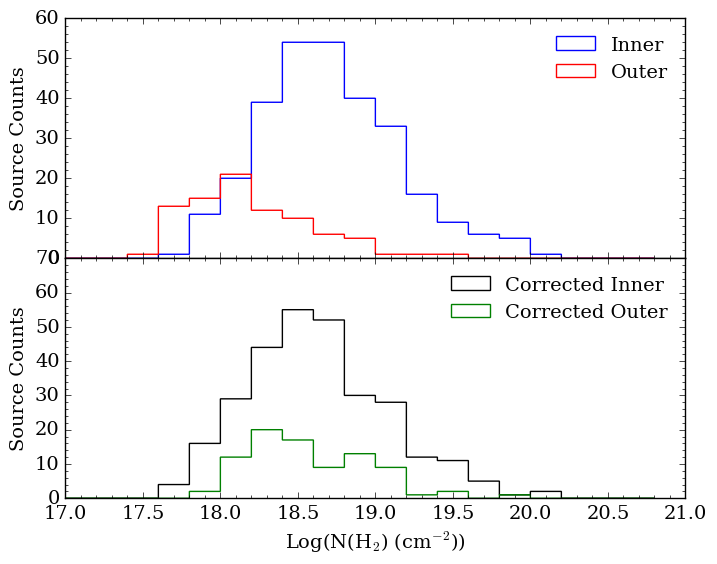}
    \caption{Upper Panel: Column densities calculated with $R=100$. Lower Panel: Distribution of sources with corrected values of $R$. Bin size is 0.2\,dex. }
    \label{fig:col_density_histo}
\end{figure}

Column densities are derived from the peak submillimetre flux of the clumps using the definition

\begin{equation}
\Big(\frac{N_{\rm{H_{2}}}}{\rm{cm}^{-2}}\Big)=\Big(\frac{S_{\nu}}{\rm{mJy}}\Big)\frac{\rm{R}}{B_{\nu}(T_{\rm{dust}})\Omega\kappa_{\nu}\mu \rm{m}_{\rm{H}}}
\end{equation}

\noindent where $\Omega$ is the beam solid angle, $\mu$ is the mean molecular weight of the gas which we assume to be 2.8 \citep{Kauffmann2008}, $\rm{m}_{\rm{H}}$ is the mass of a hydrogen atom, and $\kappa_{\nu}$ and $R$ are as defined previously. Once more, we assume $T_{\rm{dust}}=27$\,K. The distribution of column densities is shown in Figure~\ref{fig:col_density_histo} where the upper panel uses $R=100$ and the lower panel shows the values obtained using the Gianetti corrections. 

Both the inner and outer Galaxy subsets have an approximate range of $\sim$17.5-20\,dex. After the $R$ corrections, the inner Galaxy peak shifts from 18.75 to 18\,dex, with the outer Galaxy peak also moving to $\sim$18.5\,dex from 18\,dex. The inner Galaxy pre-correction values for the median and mean column densities are 18.68\,dex and 18.71\,dex while the outer Galaxy has corresponding stats of 18.12\,dex and 18.19\,dex. After the corrections, the inner Galaxy median and mean become 18.58\,dex and 18.63\,dex with the outer Galaxy similarly shifting to 18.49\, and 18.57\,dex. A KS-test gives a p-value $\ll0.001$ allowing us to reject the null hypothesis that the two samples are drawn from the same parent population.

\begin{table*}
\centering
\caption{Derived H\,{\tiny II} region properties for a selection of the full sample. Columns are as follows - Radio Name (in galactic coordinates); Submillimetre Name (as previously defined); Offset between radio and submillimetre centres; Radio Peak and Integrated fluxes; Angular Diameter; Physical Diameter; Frequency of observations; Lyman continuum flux; Bolometric Luminosity; Associated Ratios as discussed in Section 6. Full table will be available in electronic form at CDS.}
\resizebox{\textwidth}{!}{
\label{table2}
\begin{tabular}{llcccccccccccc}
\hline
\hline
{Radio Name} & {Submm Name} & {Offset} & {Peak Flux} & {Int Flux} & {d} & {d}  & {Freq} & {Log N$_{\rm{Ly}}$} & {Log L$_{\rm{bol}}$} & {Log} & {Log} & {Log} & {Log}\\
{ } & { } & {[\arcsec]} & {[mJy\,beam$^{-1}$]} & {[mJy]} & {[\arcsec]} & {[pc]}  & {[GHz]} & {[photons\,s$^{-1}$]} & {[L$_{\odot}$]} & {L$_{\rm{bol}}$/M} & {L$_{\rm{bol}}$/{M$_{\rm{corr}}$}} & {N$_{\rm{Ly}}$/M} & {N$_{\rm{Ly}}$/M$_{\rm{corr}}$} \\
\hline
$^{1}$G010.3009$-$00.1477	&	AGAL010.299$-$00.147		&	5.2	&	56.37	&	631.39	&	5.2	&	0.09	&	5	&	47.84	&	5.17	&	1.79	&	1.92	&	44.46	&	44.59	\\
$^{1}$G010.3204$-$00.2586	&	AGAL010.321$-$00.257		&	4.1	&	14.61	&	18.2	&	$-$	&	$-$	&	5	&	46.17	&	$-$	&	$-$	&	$-$	&	43.48	&	43.57	\\
$^{1}$G010.4736+00.0274	&	AGAL010.472+00.027		&	3.9	&	12.3	&	19.3	&	1.4	&	0.06	&	5	&	47.1	&	5.65	&	1.29	&	1.71	&	42.73	&	43.16	\\
$^{1}$G010.4724+00.0275	&	AGAL010.472+00.027		&	0.8	&	22.34	&	38.43	&	1.7	&	0.07	&	5	&	47.4	&	5.65	&	1.29	&	1.71	&	43.03	&	43.45	\\
$^{1}$G010.6240$-$00.3813	&	AGAL010.624$-$00.384		&	10.6	&	38.23	&	71.65	&	1.4	&	0.03	&	5	&	47.21	&	5.72	&	1.69	&	1.94	&	43.18	&	43.43	\\
$^{1}$G010.6218$-$00.3848	&	AGAL010.624$-$00.384		&	8.5	&	16.81	&	37.06	&	1.7	&	0.04	&	5	&	46.92	&	5.72	&	1.69	&	1.94	&	42.9	&	43.14	\\
$^{2}$G031.0709+00.0508	&	AGAL031.071+00.049		&	4.52	&	8.3	&	248.6	&	$-$	&	$-$	&	5	&	47.21	&	3.76	&	1.9	&	1.92	&	45.35	&	45.37	\\
$^{1}$G031.1596+00.0448	&	AGAL031.158+00.047		&	12.4	&	18.7	&	23.83	&	$-$	&	$-$	&	5	&	46.19	&	3.27	&	0.9	&	0.92	&	43.82	&	43.84	\\
$^{1}$G031.1590+00.0465	&	AGAL031.158+00.047		&	6.4	&	3.89	&	7.04	&	1.4	&	0.02	&	5	&	45.66	&	3.27	&	0.9	&	0.92	&	43.29	&	43.31	\\
$^{1}$G031.2435$-$00.1103	&	AGAL031.243$-$00.111		&	4.1	&	133.77	&	353.06	&	2.2	&	0.14	&	5	&	48.72	&	5.28	&	1.6	&	1.53	&	45.04	&	44.97	\\
$^{2}$G302.0218+00.2539	&	AGAL302.021+00.251		&	11.9	&	47.89	&	59.48	&	0.81	&	0.02	&	4.8	&	46.99	&	3.92	&	1.2	&	1.14	&	44.27	&	44.22	\\
$^{2}$G302.1525$-$00.9485	&	AGAL302.149$-$00.949		&	11.73	&	5.66	&	39.17	&	6.52	&	0.35	&	4.8	&	47.64	&	4.52	&	1.41	&	1.13	&	44.52	&	44.24	\\
$^{2}$G302.4868$-$00.0315	&	AGAL302.486$-$00.031		&	4.42	&	10.18	&	23.41	&	2.67	&	0.04	&	4.8	&	46.33	&	3.73	&	1.21	&	1.15	&	43.81	&	43.74	\\
$^{2}$G303.1173$-$00.9714	&	AGAL303.118$-$00.972		&	3.19	&	192.9	&	411.8	&	2.45	&	0.02	&	4.8	&	46.98	&	3.03	&	1.62	&	1.52	&	45.57	&	45.47	\\
$^{2}$G303.5351$-$00.5971	&	AGAL303.536$-$00.597		&	3.08	&	5.04	&	8.68	&	2.05	&	0.1	&	4.8	&	46.93	&	4.03	&	0.82	&	0.59	&	43.72	&	43.49	\\
$^{2}$G303.9976+00.2801	&	AGAL303.999+00.279		&	6.85	&	6.56	&	33.04	&	4.97	&	0.28	&	4.8	&	47.6	&	4.23	&	1.35	&	1.06	&	44.73	&	44.43	\\
$^{3}$G064.1528+01.2817	&	JCMTLSY J194815.31+280727.4		&	5.03	&	12.6	&	25.7	&	$-$	&	$-$	&	5	&	47.5	&	4.3	&	1.9	&	1.5	&	45.1	&	44.7	\\
$^{3}$G111.1919$-$00.7965	&	JCMTLSY J231545.69+595239.3		&	7.05	&	$-$	&	1.1	&	$-$	&	$-$	&	5	&	45.26	&	3.76	&	2.2	&	1.83	&	43.7	&	43.33	\\
$^{3}$G111.2824$-$00.6639B	&	JCMTLSY J231603.89+600153.8		&	4.47	&	6.5	&	90.8	&	$-$	&	$-$	&	5	&	47	&	4.37	&	1.38	&	1.06	&	44.01	&	43.69	\\
\hline
\end{tabular}
}
\end{table*}

\section{HII Region Properties}

\begin{figure}
	\subfigure{\includegraphics[width=\columnwidth]{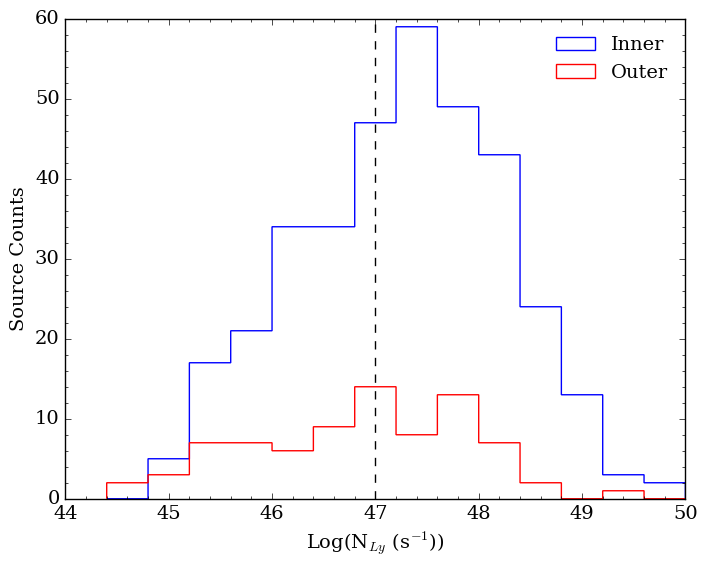}}
    \caption{Lyman continuum flux distribution for inner (blue) and outer Galaxy H\,{\tiny II} regions. The VLA completeness limit at $10^{47}$ photons\,s$^{-1}$ is shown as the vertical dashed line. The outer Galaxy sources show a much flatter distribution when compared to the inner Galaxy subset which peaks near the completeness limit. Bin size is 0.4\,dex.}
    \label{fig:LymanFlux_histo}
\end{figure}

We now examine the properties of the embedded UC H\,{\tiny II} regions using radio measurements obtained from CORNISH or the RMS database. Once more, we present the number distributions of these parameters with regards to inner and outer Galaxy subsets and remove sources within $\pm1$\,kpc of the solar circle boundary defined at $R_{0}=8.5$\,kpc. A selection of UC H\,{\tiny II} regions is given in Table \ref{table2}. The full catalogue is available online. 

\subsection{Lyman Continuum Flux}

For a massive star, one can derive the Lyman continuum flux from the radio continuum using the following equation originally presented by \cite{Carpenter1990}. 
\begin{equation}
\Big(\frac{N_{\rm{H_{2}}}}{\rm{photon}\,\rm{s}^{-1}}\Big)=9 \times 10^{43}\Big(\frac{S_{\nu}}{\rm{mJy}}\Big)\Big(\frac{D}{\rm{kpc}}\Big)^{2}\Big(\frac{\nu}{5\,\rm{GHz}}\Big)^{0.1}
\end{equation}

\noindent where $S_{\nu}$ is the integrated radio flux measured at a frequency $\nu$ and $D$ is the heliocentric distance. We assume that the UC H\,{\tiny II} regions are optically thin and that the calculations will significantly underestimate the Lyman continuum flux for more compact H\,{\tiny II} regions which will be optically thick at $\nu=5$\,GHz. The typical Lyman continuum flux error is $\sim$20\%, taking into account the 10\% errors in both distance and the average flux measurement from RMS.  

Figure~\ref{fig:LymanFlux_histo} shows the distribution of fluxes for the 350 inner (blue) and 81 outer (red) Galaxy sources. Radio observations for both CORNISH and RMS data were obtained with either VLA or ATCA, giving a consistent completeness limit of $\sim10^{47}$\,photons\,s$^{-1}$. Both subsets have a similar range and peak near this value, although the outer Galaxy sources have a flatter distribution. The medians for both subsets also fall near the limit at 47.29\,dex for the inner Galaxy and 46.95\,dex for the outer Galaxy as do the mean values of 47.24\,dex and 46.79\,dex, respectively. A KS-test yields a p-value of 0.017. In this case, we are unable to reject the null hypothesis that the H\,{\tiny II} regions are derived from a different parent distribution. This fact will be relevant for further discussion in Section 6. 

\subsection{HII Region Size}

\begin{figure}
   	\includegraphics[width=\columnwidth]{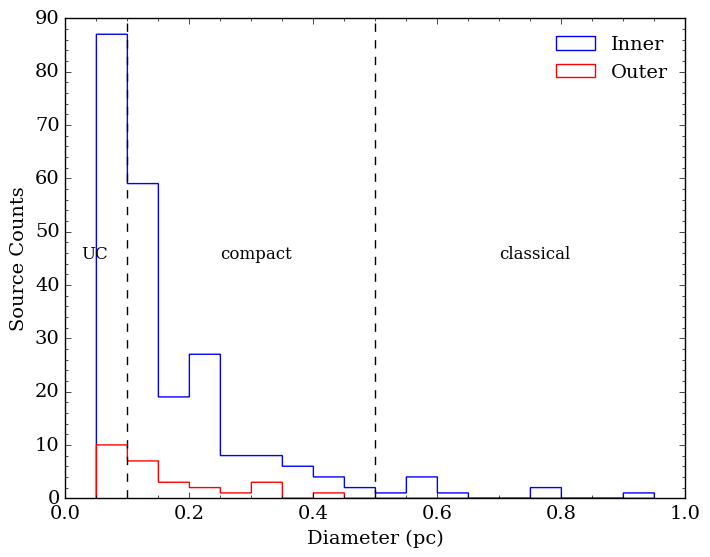}
      \caption{Distribution for diameter sizes of H\,{\tiny II} regions. Vertical lines represent the cut-offs for UC, compact, and classical H\,{\tiny II} regions normally used in the literature. Bin size is 0.05\,pc.}
    \label{fig:HIIsizes_histo}
\end{figure} 

Physical sizes for the UC H\,{\tiny II} regions of each subset are given in Figure~\ref{fig:HIIsizes_histo}. The outer Galaxy sample has a much smaller range falling between 0.1 and 0.5\,pc in diameter, while inner Galaxy sources have an upper limit closer to 1\,pc. Nonetheless, both samples show most H\,{\tiny II} regions falling within the compact and UC limits, as expected. The dashed vertical lines represent the arbitrary size limits usually assigned to UC, compact, and classical H\,{\tiny II} regions. Both inner and outer Galaxy sources have a mean and median value around 0.1\.pc. The inner Galaxy outliers with $d > 0.5$\,pc appear to be objects that had the highest angular sizes and also fell near to the distance error cutoff limits around R$_{\rm{GC}}=8.5\pm1$\,kpc. These objects will suffer the most from the associated distance uncertainties, thus skewing calculations for the physical diameter.  All sources fall well below the maximum accepted 1\,pc limit and may be regarded as `compact' sources. Associated uncertainties will be dominated by the distance errors. A KS-test for this property does allow us to reject the null hypothesis giving a p-value result of $p\ll0.001$.

\begin{figure}
	\includegraphics[width=\columnwidth]{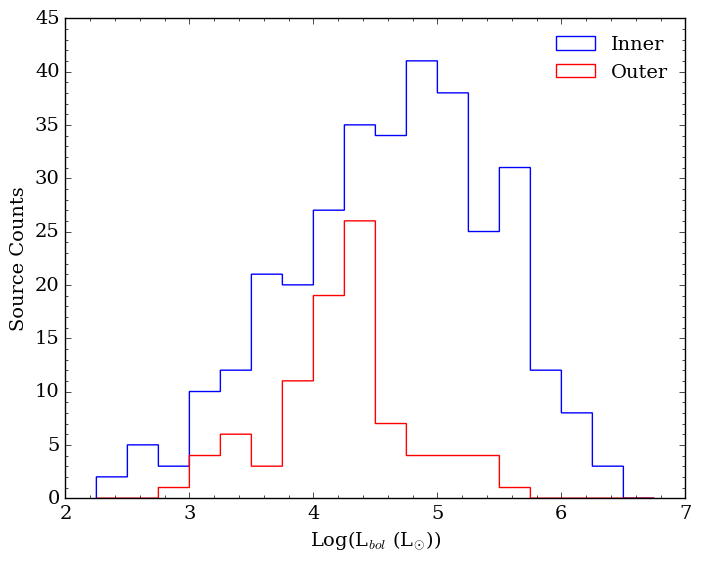}
    \caption{Bolometric luminosities for both subsets were adopted from the RMS catalogue. See text for further details. Bin size is 0.25\,dex.}
    \label{fig:Lbol_histo}
\end{figure}

\subsection{Bolometric Luminosity}

\begin{figure*}
	\includegraphics[width=\textwidth]{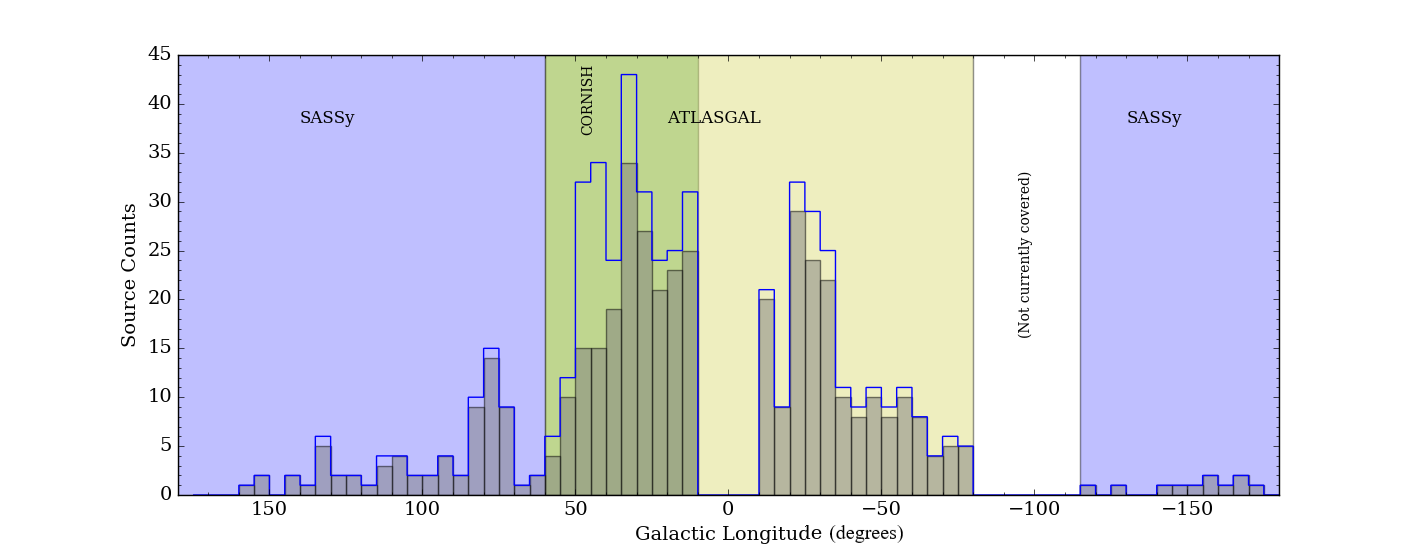}
    \caption{The distribution in Galactic longitude for the complete sample. Host clumps are shown as the filled histogram with H\,{\tiny II} regions as the blue line. The background has been shaded according to the submillimetre survey which studied the corresponding longitudinal range. CORNISH has also been highlighted to illustrate the scope of the original catalogue. The inclusion of the rest of ATLASGAL and the more recent SASSy data greatly increase the area covered. Peaks can be observed where particularly intense regions of star formation have a high multiplicity of H\,{\tiny II} regions per host clump. Bin size is 5 degrees. }
    \label{fig:gal_long_histo}
\end{figure*}

The use of RMS allows us to include the estimated bolometric luminosities of each source \citep{Mottram2011a, Mottram2011b}. The RMS survey is complete to a few $10^{4}$ L$_{\odot}$ for the population of mid-infrared-selected H\,{\tiny II} regions and MYSOs sample. \citealt{Urquhart2013cornish} noted that this made it well matched to comparisons with CORNISH and we extend that comparison with our additional collection of H\,{\tiny II} regions. 

In Figure \ref{fig:Lbol_histo}, the inner Galaxy sources cover a wider range within $\sim$2-6.5\,dex while the outer Galaxy subset is more limited between 3-6\,dex. These limits equate to a B3 star and an O3 star, respectively \citep{Martins2005}. The inner Galaxy sample peaks at 5\,dex but the outer Galaxy peak is closer to $\sim$4.3\,dex. The median and average luminosity values for the inner Galaxy are identified to both be around 4.6\,dex while the outer Galaxy values fall near 4.3\,dex. The same trend appears for average and median values with outer Galaxy sources having average luminosities of 4.22\,dex vs. the inner Galaxy's 4.62\,dex and an outer Galaxy median at 4.26 vs. the inner Galaxy median of 4.73\,dex. A KS-test conclusively rejects the null hypothesis with $p\ll0.001$. The fluxes used to estimate the luminosities are effectively clump-averaged values and thus, are a measure of the total luminosity of the embedded YSOs. In reality, it will not be a single star embedded within the H\,{\tiny II} region but several with the most massive star being a few subclasses later than expected. This agrees with the results found in Section 4 where the inner and outer Galaxy clumps appeared to most likely be distinctly different samples. Typical errors for these values are approximately 34$\%$ \citep{Mottram2011b}. 

\section{Discussion}

The supplementary UC H\,{\tiny II} regions have more than doubled the sample size of the original ATLASGAL-CORNISH study. With an additional 323 H\,{\tiny II} regions and 275 host clumps, the total UC H\,{\tiny II} region catalogue now includes 536 UC H\,{\tiny II} regions associated with 445 molecular clumps. These cover a wide range of Galactic longitudes encompassing $\sim$80\% of the Galactic plane. The farthest object is nearly 20\,kpc away from the Sun and we have extended the sample to include sources out to $R_{\rm{GC}}\sim15$\,kpc, allowing us to examine any trends in massive star formation across a more complete scope of Galactic plane variation.

\subsection{Galactic Distribution}

\begin{figure}
	\includegraphics[width=\columnwidth]{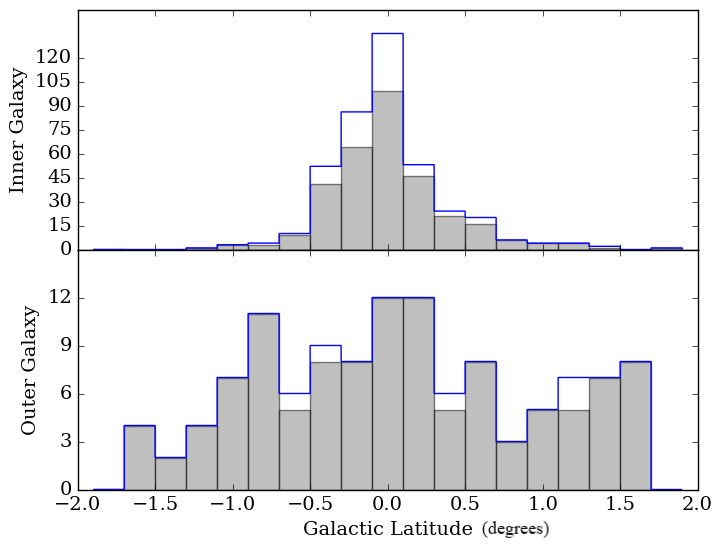}
    \caption{Distribution for star forming clumps (grey filled) and embedded H\,{\tiny II} regions (blue line) of the complete sample. The upper panel illustrates the distribution for sources with $R_{\rm{GC}} < 8.5$ and the lower panel shows the same for sources with $R_{\rm{GC}} > 8.5$. The broader distribution of the outer Galaxy is in part due to the flare in the molecular disk with increasing radius also from projection, with closer sources such as those found in SASSy having higher observed angles than the ATLASGAL sources observed at much farther distances. Bin size is 0.2 degrees.}
    \label{fig:gal_lat_histo}
\end{figure}

Figure~\ref{fig:gal_long_histo} shows the Galactic longitude distribution of the full catalogue. Host clumps are plotted as the filled grey histogram and the embedded H\,{\tiny II} regions are shown as the blue line. The addition of the remaining ATLASGAL and SASSy regions allows the total sample to cover the majority of the Galactic plane with the exceptions of  $-120\degr \leq l \leq -80\degr$ for which there are currently no equivalent submillimetre observations, and the Galactic centre ($-10\degr \leq l \leq 10\degr$) that we have avoided due to source confusion. Peaks in the number of H\,{\tiny II} regions per clump are clearly seen, indicating areas of intense star formation. These correspond to well known star forming complexes. For example, the effects from W43, W51, and W49A are visible within $30\degr \leq l \leq 50\degr$ \citep{Eden2017, Urquhart2014agalrms}.

Figure~\ref{fig:gal_lat_histo} shows the distribution in Galactic latitude. ATLASGAL observed sources with $|b| < 1.5$\degr\space while the SASSy range included those with $|b|<2$\degr. In the inner Galaxy, star formation appears to be mostly restricted to the Galactic plane with a skew towards negative latitudes due to the Sun's position as well as from the observed warp in the Galactic plane. The outer Galaxy shows a broader distribution. This is in part due to the fact that the molecular disk flares out with galactocentric radius \citep{wouterloot1990} and also the overall closer SASSy sources will have larger ranges of latitudes observed than the more distant ATLASGAL sources.

\subsection{Galactic Trends}

\subsubsection{Clump Mass \& Lyman Continuum Flux}

\begin{figure}
    \includegraphics[width=\columnwidth]{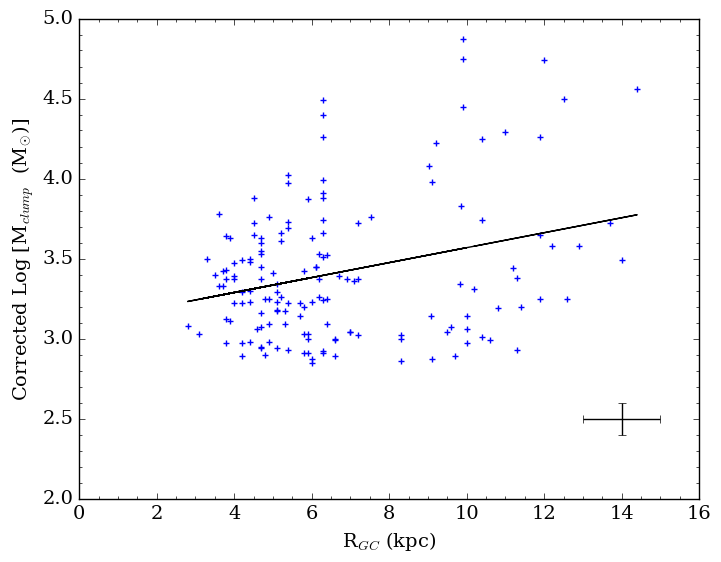}
    \caption{Corrected masses of host star forming clumps as a function of Galactocentric radius. We have applied a uniform mass cutoff limit of $10^{2.85}$\,M$_{\odot}$ and removed sources with $D>8$\,kpc. The vertical strip of sources near $R_{\rm{GC}}\approx6$\,kpc consists of Cygnus X sources whose distances are known to a higher degree of accuracy than the surrounding data. See text for details. }
    \label{fig:Dist-Mass}
\end{figure}

\begin{figure}
    \includegraphics[width=\columnwidth]{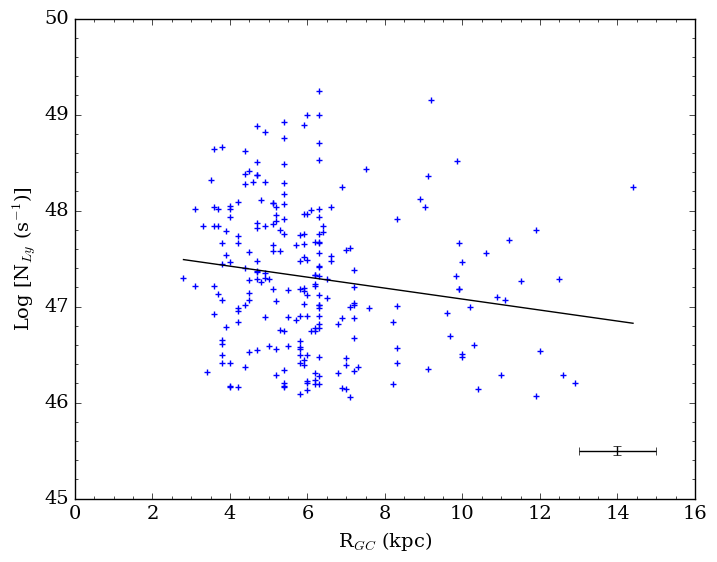}
    \caption{The relationship of Lyman continuum flux to galactocentric radius of UC H\,{\tiny II} regions above the flux completeness limit of $10^{46.06}$\,photons s$^{-1}$ and with $D>8$\,kpc. }
    \label{fig:Distance_LymanFlux}
\end{figure}

In Figure~\ref{fig:Dist-Mass}, we examine the relationship between clump mass and Galactocentric radius. The estimation of clump masses relies heavily on accurate distance measurements and we must be cautious of the completeness limits associated with each survey to ensure that any trends we may observe are not a result of the Malmquist bias. ATLASGAL observed across the inner Galaxy mid-plane covering $280\degr \leq l \leq60\degr$, while SASSy was directed away from the centre and examined the overall closer regions of the outermost edges of the plane which include $60\degr \leq l \leq 240\degr$ (see Figure~\ref{fig:survey_coverage}). We determine their respective sensitivity limits by considering a hypothetical source at a distance of 8\,kpc. At this range, ATLASGAL is complete for all inner Galaxy sources ($R_{\rm{GC}}\approx2-8$\,kpc) and SASSy is complete for the outer radii ($R_{\rm{GC}}\approx8-15$\,kpc). Using a significance level of 5$\sigma$ and adopting the average rms values for each survey, we calculated mass completeness limits of $10^{2.85}$\,M$_{\odot}$ and $10^{2.48}$\,M$_{\odot}$ for ATLASGAL and SASSY, respectively. For consistency, we use the higher of these values and remove any sources that fall below $10^{2.85}$\,M$_{\odot}$. We also remove sources with heliocentric distance $D > 8\,$kpc since we will not be able to detect a complete distribution in masses beyond this boundary. In the end, we are left with 154 clumps, ranging from $\sim10^{3}$ to $10^{5}$\,M$_{\odot}$ (using corrections from \citealt{Giannetti2015} as previously discussed) across $R_{\rm{GC}}=2-15$\,kpc. Representative error bars are given in the lower righthand corner of the Figure~\ref{fig:Dist-Mass}. 

We also calculated the equivalent completeness limits for the Lyman continuum fluxes. Once more, we consider a source at the maximum distance (8\,kpc) and use the mean rms values as found in the RMS radio follow-up and CORNISH survey papers (\citealt{Urquhart2007}; \citealt{Urquhart2009}; \citealt{Purcell2013}). The same 5$\sigma$ significance level is applied. We find Lyman flux cutoffs at $10^{45.93}$ and $10^{46.06}$ photons s$^{-1}$ for RMS and CORNISH data, respectively. We again implement the higher limit of $10^{46.06}$ photons s$^{-1}$ as well as the distance cutoff limit for sources with $D>8$\,kpc. The results are plotted as a function of $R_{\rm{GC}}$ in Figure~\ref{fig:Distance_LymanFlux} for 238 UC H\,{\tiny II} regions.

Both Figures~\ref{fig:Dist-Mass} and~\ref{fig:Distance_LymanFlux} contain a great deal of scatter in the data, particular with sources in the outer Galaxy. In the case of Lyman flux, the inner Galaxy sources are also significantly scattered. To better interpret the data, we include trendlines representing the unweighted least squares linear regression fits for each dataset. In Figure~\ref{fig:Dist-Mass}, clump mass appears to increase with increasing $R_{\rm{GC}}$ but has a relatively flat slope of $+0.05\pm0.01$ kpc$^{-1}$. We recall that our use of a single average dust temperature for all sources (T$_{\rm{dust}}=27 $\,K, see Section 3.2) may result in overestimating clump masses at higher radii. Indeed, a Spearman rank test gives a correlation of 0.123 with a high p-value of 0.11. We determine that the recorded slope is consistent with being flat and conclude that there is no correlation between clump mass with Galactocentric radius. In Figure~\ref{fig:Distance_LymanFlux}, we observe an overall decrease in Lyman flux but with another shallow slope of $-0.06\pm0.02$ kpc$^{-1}$ and there is a large amount of scatter present. The scatter is likely a result from using the optically thin assumption for all H\,{\tiny II} regions (see Section 5.1) which would underestimate the Lyman continuum flux for the most compact regions that are optically thick at $\nu=5$\,GHz. A Spearman test gives a correlation of $-0.078$ with another high p-value of 0.22. We conclude that there is also no significant correlation between Lyman continuum flux with Galactocentric radius. In the next step, we will examine the  distance-independent variables of $N_{\rm{Ly}}/M$ and $L_{\rm{bol}}/M$ and how their corresponding trends correlate with previous studies.

\subsubsection{A Proxy for Star Formation Efficiency \& Related Quantities}

By examining the Lyman flux and the luminosity of the H\,{\tiny II} regions per unit mass, we are able to remove the distance dependence. As such, $N_{\rm{Ly}}/M$ will serve as a proxy for the massive star formation efficiency while $L_{\rm{bol}}/M$ will represent the overall star formation efficiency in massive star forming clumps but includes both the low and high mass stars. This gives us the added benefit of being able to study the whole catalogue as we no longer have to eliminate sources due to completeness limits. 

\begin{figure}
    \includegraphics[width=\columnwidth]{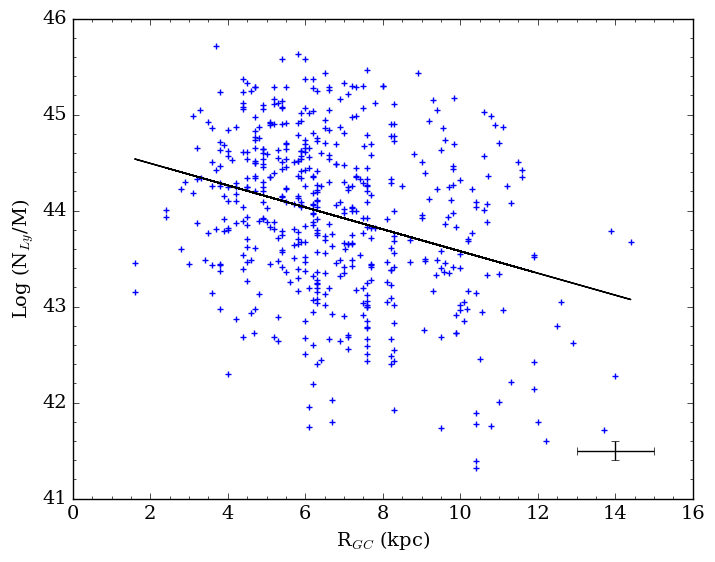}
    \caption{The Lyman continuum flux or rate of UV-photon output per unit mass serves as a proxy for the star formation efficiency of the most massive stars. We observe a downward trend when plotted again galactocentric radius but the scatter presents a high degree of uncertainty. See text for details. }
    \label{fig:Distance_Lyman-Mass}
\end{figure}

\begin{figure}
      \includegraphics[width=\columnwidth]{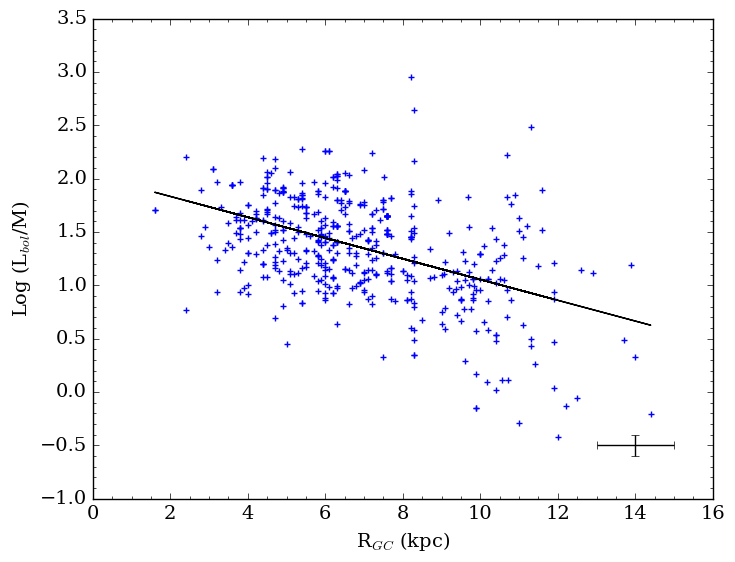}
    \caption{The relation $L_{\rm{bol}}/M$ serves as a proxy for overall star formation efficiency plotted against galactocentric radius. Agsain, we observe a downward trend where star formation appears to be less efficient in the outer Galaxy than in the inner Galaxy. This agrees with Figure~\ref{fig:Distance_Lyman-Mass} which shows the trend for star formation efficiency of only the massive stars.}
    \label{fig:distance_Lbol-M}
\end{figure}

These distance-independent parameters are shown in Figures~\ref{fig:Distance_Lyman-Mass} and~\ref{fig:distance_Lbol-M}. Both plots show a downward trend, initially suggesting that the massive and overall star formation efficiencies are lower in the outer Galaxy than in the inner Galaxy. However, Figure~\ref{fig:Distance_Lyman-Mass} has a large amount of scatter and this affects the significance of the result. For the 518 H\,{\tiny II} regions plotted, we estimate a gradient of $-0.11\pm$0.02 kpc$^{-1}$, but a Spearman test gives a correlation number of -0.075 with p-value of 0.24 which remains well above our desired 3$\sigma$ confidence level and thus shows no significant correlation.  Figure~\ref{fig:distance_Lbol-M} yields a slope of $-0.1\pm0.008$ kpc$^{-1}$ and this decline over $R_{\rm{GC}}$ can be confirmed with the Spearman rank test giving a correlation of $-0.258$ with a p-value $\ll0.001$. Thus, we conclude that the overall star formation efficiency for massive star forming clumps is decreasing and is lower in the outer Galaxy.

While there has been little previous investigation into these exact quantities as a function of Galactocentric radius, \citealt{Lada2010} previously showed that star formation was observed to be closely correlated with dense gas. The decline in overall star formation efficiency seen in Figure~\ref{fig:distance_Lbol-M} is mirrored in the results by \citealt{RomanDuval2016} who traced the ratio of dense ($^{13}$CO-emitting) gas surface density to the total ($^{12}$CO + $^{13}$CO) gas surface density and found an average gradient of $-0.06$\,kpc$^{-1}$ over $3 \leq R_{\rm{GC}} \leq 8$\,kpc. A decreasing trend is also seen in \citealt{Nakanishi2006} who plotted the molecular gas surface density as a function of galactocentric radius and found a noticeable decline, dropping steeply just beyond the solar neighbourhood or the start of the outer Galaxy by our previous definition. The same was observed in the nearby galaxy NGC 6946 by \citealt{Schruba2011}. \citealt{Misiriotis2006} presented an estimate of the surface density of the star formation rate which declined steeply past $R_{\rm{GC}}=5$\,kpc though the data was acquired from older COBE observations of extended H\,{\tiny II} regions \citep{Boggess1992}. Finally, \citealt{Ragan2016} defined a star forming fraction (SFF) as the ratio of Hi-GAL objects associated with a 70\,$\mu$m component with the total catalogue of Hi-GAL sources. The SFF may be considered as the fraction of dense clumps with embedded YSOs. They derived a mean SFF in the Galactic disk ($3.1 < R_{\rm{GC}} < 8.6$\,kpc) of 25\% that declined with $R_{\rm{GC}}$ at a rate of $-0.026\pm0.002$\,kpc$^{-1}$.  However, a few other studies have found no change in the dense gas between the inner and outer Galaxy, including \citealt{Battisti2014} who defined a dense gas mass fraction as the ratio of the mass traced by submillimetre dust emission to the mass of the parent cloud traced by $^{13}$CO. They found no dependence on $R_{\rm{GC}}$ but were limited to a range over $3 \leq R_{\rm{GC}} \leq 8$\,kpc. \citealt{Eden2013} also observed the ratio between clump to cloud mass to determine a clump formation efficiency and found no difference between the inter-arm and spiral-arm region along the selected line of sight ($37.83\degr \leq l \leq 42.50\degr$) which translates to a galactocentric range of $4 \leq R_{\rm{GC}} \leq 8.5$\,kpc. It must be noted that all of these studies are primarily restricted to the inner Galaxy and show very subtle trends which are only apparent with large samples (e.g. \citealt{RomanDuval2016}) or large $R_{\rm{GC}}$ range.

\begin{figure}
    \includegraphics[width=\columnwidth]{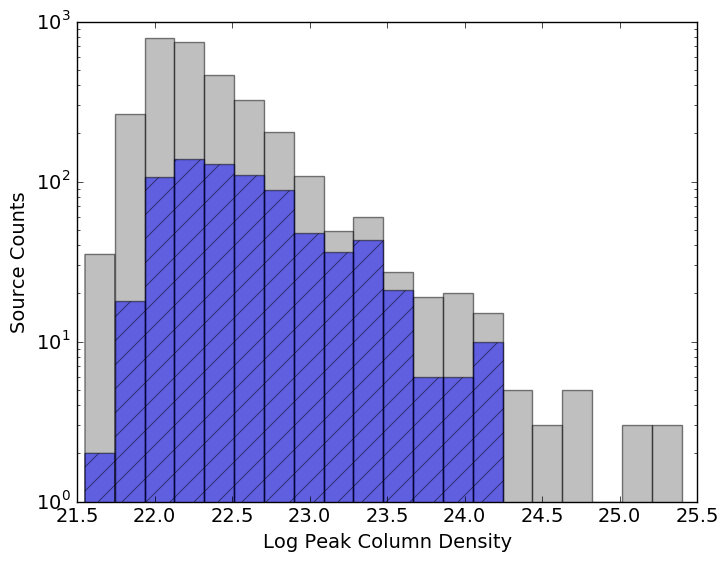}
    \caption{Column densities of the full SASSy submillimetre clump catalogue (grey). Previously observed clumps with known tracers of massive star formation are overlaid in blue. There are a relatively small number ($\sim$20) of SASSy clumps with column density greater than $10^{23}$\,cm$^{-2}$ and no known presence of massive star formation. Bin size is 0.2 dex.}
    \label{fig:vla_col_den}
\end{figure}

As noted in Section 3, we must address the possibility of whether the low number of matches between the SASSy OGS and RMS data may result in an incomplete catalogue for higher values of galactocentric radius, thus skewing our conclusions. \citealt{Urquhart2014agalrms} showed that the massive star-forming clumps are strongly associated with the highest column-density clumps in the Galaxy. Nearly all clumps with column density above 2-3 x $10^{23}$\,cm$^{-2}$ are host to massive star formation \citep{Urquhart2014agalrms}. In Figure~\ref{fig:vla_col_den} we show the column density distribution for all SASSy submillimetre sources (grey) along with the distribution of those particular clumps that are known to play host to massive star formation based on a review of the current literature. There are $\sim$20 SASSy clumps above the crucial column density limit that do not show associated tracers of massive star formation. Compared to the 445 clumps that we currently have in our final catalogue, this value represents less than 5\% of the total number of Galactic clumps and thus its absence will not have a significant effect on the trends we derive.

Overall, we see downward trends with increasing $R_{\rm{GC}}$ for both the amount of massive stars per unit clump mass as well as overall stars per unit clump mass (in the massive star forming clumps) although only the latter is confirmed as a significant result. This drop in the overall star formation efficiency combined with a flat trend in the massive star formation efficiency may imply a shift in the initial mass function (IMF) towards a lesser fraction of lower mass stars in the outer Galaxy. However, we are unable to draw any robust conclusions because we only see a statistically significant trend in the overall star formation efficiency or L$_{\rm{bol}}/$M. The remaining quantities contain too much scatter to confidently identify any trends and would require further investigation in order to determine the effects if any on the IMF.

\section{Summary \& Conclusions}

In summary, we have used the methods of \citealt{Urquhart2013cornish} and \citealt{Urquhart2014agalrms} and incorporated outer Galaxy data from SASSy to compile a sample of compact and UC H\,{\tiny II} regions that covers the Galactic plane from $280\degr \leq l \leq 240\degr$. These massive star forming regions have been identified via cross-matching and coincidence of infrared, radio, and submillimetre data with ATLASGAL providing the bulk of inner Galaxy molecular clumps while SASSy encompassed the outer Galaxy. The complete catalogue totals 536 UC H\,{\tiny II} regions associated with 445 host clumps. From this catalogue, we are able to draw the following conclusions:  
\begin{itemize}
    \item We have compiled or calculated the host clump properties and find that the inner Galaxy clumps tend to be similar to their outer Galaxy counterparts as shown by a 2-sample Kolmogorov-Smirnov test.
    \item However, by the same test, there appears to be some difference in how star formation is happening and the inner and outer Galaxy H\,{\tiny II} regions are clearly drawn from two different parent samples.
    \item We  plotted the clump mass and Lyman continuum flux as a function of galactocentric radius finding that mass tended to increase with increasing $R_{\rm{GC}}$ while Lyman flux decreased, although the results could not be proven as significant due to the large degree of scatter and error in the data.
    \item We removed the distance errors by also plotting the massive star formation efficiency or number of massive stars per unit mass (N$_{\rm{Ly}}/$M) and the overall star formation efficiency or total number of stars per unit mass (L${\rm{bol}}/$M) as a function of $R_{\rm{GC}}$.
    \item By the Spearman rank test, we confirmed a significant correlation for L$_{\rm{bol}}/$M with Galactocentric radius.
    \item We conclude that the overall star formation efficiency is lower in the outer Galaxy than in the inner Galaxy
\end{itemize}

\section*{Acknowledgements}

SASSy data was acquired via the James Clerk Maxwell Telescope (JCMT). The JCMT is operated by the East Asian Observatory on behalf of The National Astronomical Observatory of Japan, Academia Sinica Institute of Astronomy and Astrophysics, the Korea Astronomy and Space Science Institute, the National Astronomical Observatories of China and the Chinese Academy of Sciences (Grant No. XDB09000000), with additional funding support from the Science and Technology Facilities Council of the United Kingdom and participating universities in the United Kingdom and Canada. We acknowledge support from the United Kingdom Science \& Technology Facility Council via grants ST/M001008/1 and ST/R000905/1. We also extend thanks to the valuable input from Toby Moore and David Eden in the final stages of the discussion. We acknowledge previous work and assistance provided by Gaius Manser in the early stages of the SASSy catalogue and matching, primarily with the SASPER dataset. Finally, we wish to thank the anonymous reviewer for valuable insight into clarifying certain details. This work would not have been possible without access to NASA ADS.

%The Acknowledgements section is not numbered. Here you can thank helpful
%colleagues, acknowledge funding agencies, telescopes and facilities used etc.

%%%%%%%%%%%%%%%%%%%%%%%%%%%%%%%%%%%%%%%%%%%%%%%%%%



\bibliographystyle{mnras}
\bibliography{Mendeley} % if your bibtex file is called example.bib
%\end{document}

%\begin{thebibliography}{99}
%\end{thebibliography}

%%%%%%%%%%%%%%%%%%%%%%%%%%%%%%%%%%%%%%%%%%%%%%%%%%

%%%%%%%%%%%%%%%%% APPENDICES %%%%%%%%%%%%%%%%%%%%%

%%%%%%%%%%%%%%%%%%%%%%%%%%%%%%%%%%%%%%%%%%%%%%%%%%

% Don't change these lines
\bsp	% typesetting comment
\label{lastpage}
\end{document}